\documentclass[submission, Phys]{SciPost}
\usepackage{authblk,amsmath,amssymb,amsthm,cite,epsfig,url,psfrag,eepic,mathtools,mathtools,graphicx,xcolor,rotating,mathrsfs}
\usepackage[normalem]{ulem}
\newtheorem{theorem}{Theorem}[section]

\newtheorem{lemma}[theorem]{Lemma}
\usepackage{simplewick}
\usepackage{hyperref}
\usepackage{graphicx}
\usepackage{float}
\usepackage{epstopdf}
\usepackage[process=none]{pstool}

\makeatletter
\def\Appendix{\appendix
	\def\@seccntformat##1{Appendix~\csname the##1\endcsname.~~}}
\makeatother
\makeatletter
\newcommand*{\rom}[1]{\expandafter\@slowromancap\romannumeral #1@}
\title{Properties of the temporal transfer matrix in integrable Floquet circuits}
\makeatletter
\def\Appendix{\appendix
	\def\@seccntformat##1{Appendix~\csname the##1\endcsname.~~}}
\makeatother
\makeatletter
\@addtoreset{equation}{section}

\makeatother

\makeatletter
\newcommand{\zerounderset}[3][\mathord]{%
	#1{\vtop{
			\let\\\cr
			\baselineskip\z@skip\lineskip.25ex
			\ialign{\hidewidth$##$\hidewidth\crcr
				\omit$#3$\cr
				#2\crcr
			}%
	}}%
}
\makeatother

\begin{document}
\author[1,2]{Ilya Vilkoviskiy}
\affil[1]{Department of Theoretical Physics, University of Geneva, Quai Ernest-Ansermet 30, 1205 Geneva, Switzerland}
\author[3]{Kirill Matirko}
\affil[2]{Department of Physics, Princeton University, Princeton NJ 08544, USA}
\affil[3]{Department of Mathematics, HSE University, Moscow 119048, Russia}
\date{\today}
\maketitle

\begin{abstract}
One possible approach to studying non-equilibrium dynamics is the so-called influence matrix (IM) formalism. The influence matrix can be viewed as a quantum state that encodes complete information about the non-equilibrium dynamics of a boundary degree of freedom. It has been shown that the IM is the unique stationary point of the temporal transfer matrix. This transfer matrix, however, is non-diagonalizable and exhibits a non-trivial Jordan block structure.

In this article, we demonstrate that, in the case of an integrable XXZ spin chain, the temporal transfer matrix itself is integrable and can be embedded into a family of commuting operators. We further {conjecture} the exact expression for the IM as a particular limit of a Bethe wavefunction, with the corresponding Bethe roots given explicitly.

We also focus on the special case of the free-fermionic XX chain. In this setting, we uncover additional local integrals of motion, which enable us to analyze the dimensions and structure of the Jordan blocks, as well as the locality properties of the IM. Moreover, we construct a basis of quasi-local creation operators that generate the IM from the vacuum state.
\end{abstract}
\section{Introduction}

Understanding the universality classes of correlation functions in quantum systems is one of the central goals of quantum many-body physics. While ground-state properties are comparatively well understood, real-time correlations remain challenging. Integrable models provide a uniquely controlled setting. On the microscopic side, form-factor expansions yield exact expressions for dynamical correlation functions \cite{caux2005computation,kitanine2005algebraic,babenko2021exact}. In practice, however, these methods are most effective for two-point functions and become hard to extend to higher-point or long-time regimes.

At the opposite, coarse-grained end, generalized hydrodynamics (GHD) captures Euler-scale transport and post-quench evolution of local observables \cite{castro2016emergent,bertini2016transport}. GHD organizes late-time dynamics in terms of quasiparticle densities and their effective velocities, providing predictions for transport coefficients and hydrodynamic tails. Yet refined phenomena such as KPZ-type dynamical scaling \cite{KPZ2019} and the systematic computation of correlation functions in the presence of weak integrability breaking are still not fully understood, see \cite{krajnik2020kardar} for discussion.

A natural way to bridge these microscopic and hydrodynamic descriptions is through controlled real-time simulations that preserve integrability. Integrable Floquet dynamics offer precisely such a playground: they implement exact, stroboscopic time evolution within an integrable framework, enabling direct access to space-time dependent correlators, finite-density quenches, and scaling crossovers beyond the strict Euler scale \cite{DESTRI1987363,destri1995local,vanicat2018integrable,miao2024floquet}.

In this work, we introduce a framework inspired by the so-called influence matrix (IM) approach to dynamical Floquet systems, introduced in \cite{banuls2009matrix} and further developed in \cite{muller2012tensor,hastings2015connecting,lerose2021influence}. The IM is a quantum state that encodes complete information about the non-equilibrium dynamics of a boundary degree of freedom. 

{Beyond the numerical success, there are only a few models, for which the IM is known exactly. Few known examples are provided by the Rule 54 cellular automata model \cite{klobas2021exactre,klobas2021exactth} 
and also a large class of \textit{SWAP-control} circuits \cite{wang2025temporal} where the authors provided an exact MPS solution for the corresponding IM, allowing to exactly compute all possible correlation functions of operators inserted at the same spatial spin. However, all of these examples are fine-tuned, and do not represent generic features of integrable dynamics.}

{In this article, we make a step towards understanding the properties of the IM for integrable systems beyond these isolated examples}. It has been shown {in \cite{lerose2021influence}} that the IM is the unique stationary point of the temporal transfer matrix. This transfer matrix, however, is non-diagonalizable and exhibits a non-trivial Jordan block structure. 

For an evolution with $N$ Floquet time steps, we introduce a commuting family of transfer matrices $T(v)$ associated with an XXZ chain of length $4N$ with four alternating inhomogeneities. {Such that the IM itself is a joint eigenvector of all the commuting transfer matrices and can be considered as Bethe vector.} {Similar observations were made in \cite{klobas2021entanglement,klobas2024non} for the case of Rule 54 model allowing to compute the charge fluctuations and growth of entanglement. This approach was generalized to the case of generic integrable model \cite{bertini2022growth,bertini2023nonequilibrium} allowing to deduce the coarse-grained GHD description of the IM}.

As reported in Ref.~\cite{giudice2022temporal}, at least close to the vicinity of a special exactly solvable dual-unitary (DU) point (see a recent review \cite{bertini2025exactly}), the IM exhibits logarithmic growth of temporal entanglement entropy, formulating a conjecture that it might be a generic feature of the XXZ integrable system.
 
{Motivated by the sublinear growth of temporal entanglement and the structure of the IM, we undertake a microscopic study of the transfer matrix and the integrable structures underlying it.} Remarkably, we have found that the Bethe vector corresponding to the IM has very specific and explicit Bethe roots, provided by the exact solution of the Bethe ansatz equations (BAE). Our approach differs from the quantum transfer matrix (QTM) formalism introduced in \cite{gohmann2004integral}  and adapted for dynamical observables in \cite{sakai2007dynamical}. The QTM method is formulated specifically in the continuous-time limit and requires taking a careful scaling limit to recover physical dynamics. While our setup can also be tuned toward a continuous-time regime, the integrable structure we construct defines a local Floquet evolution with a finite number of time steps. This leads to novel spectral and quasiparticle properties of the temporal transfer matrix.

It is worth mentioning that similar transfer matrices were introduced in \cite{claeys2022correlations}, where the authors studied correlation functions of spin operators in brick-wall Floquet dynamics. However, in our approach, correlation functions are computed by sandwiching local observables between two IMs utilizing the logarithmic entanglement, instead of a trace calculation in \cite{claeys2022correlations} which demands the knowledge of the whole spectrum of the temporal transfer matrix. Similarly to \cite{claeys2022correlations}, we observed that the transfer matrix is non-diagonalizable. 

Analysis of an integrable system typically begins with the spectrum of its transfer matrix. In the degenerate case, the relevant characterization is instead provided by the pattern of degeneracies and the structure of the associated Jordan blocks. These features have not been systematically explored; here we take the first step toward filling this gap. The non-diagonalizability can be explained at the level of quasiparticles. We
observed massive quasiparticles related to spin-one excitations, as well as light quasiparticles with Bethe roots concentrated around one of the two inhomogeneities of the transfer matrix. The light quasiparticles are responsible for the Jordan block structure. Our results offer a fresh perspective on integrable dynamical systems. Moreover, the influence matrix (IM) machinery extends beyond correlation functions, as it can also address quantum quenches and dynamics in the presence of an integrability-breaking impurity. We hope this framework will open a practical route to study these phenomena efficiently.

In order to better understand the structure of Jordan blocks, we analyzed the free-fermionic XX limit. Remarkably, we have found the local non-Hermitian Hamiltonians that commute with the transfer matrix $T(v)$ and have the IM as a unique ground state with zero energy. Due to non-Hermiticity, these Hamiltonians share the same Jordan block structure as the transfer matrix. We note that a similar structure was studied in \cite{prosen2010spectral} in the context of the third quantization of Lindbladian dynamics, and the Jordan blocks were provided by \cite{kitahama2024jordan}. Thus, our model provides an interesting and unexpected playground to study non-Hermitian physics. In contrast to the interacting case, here we can study and fully classify the emergent Jordan structure. {The space of temporal trajectories associated with an evolution of $N$ Floquet periods admits a single-particle description. We found that the $4N$-dimensional space of single-particle wave functions decomposes into four Jordan blocks, each of size $N$.} The corresponding many-body half-filled state is obtained by occupying $2N$ quasiparticles among the $4N$ available generalized eigenstates. In this language, the IM corresponds to two Jordan blocks being fully occupied, while the remaining two are empty.

The quasiparticles we find are not local excitations and typically exhibit long-range temporal support.
Nevertheless, one can form special linear combinations that are quasi-local, with power-law tails. Such a basis can be constructed, for instance, by applying a Gram--Schmidt orthogonalization procedure to the Jordan-block basis. The existence of a quasi-local basis implies a gapped quasi-local Hamiltonian $H_{\text{gap}}$, whose tails decay as $r^{-3/2}$. This confirms an analogous observation for the kicked Ising model at an integrable point~\cite{PhysRevB.104.035137}.

{Despite the simplicity of a ground state, the evolution of generic states under temporal transfer matrix is highly non-trivial, generating highly entangled intermediate states.} This phenomenon was reported earlier and is responsible for the entanglement barrier \cite{lerose2023overcoming,yao2024temporal}. In order to address this, we provide the canonical Jordan form of the local Hamiltonians and the transfer matrix on the single-particle sector and then, following \cite{kitahama2024jordan}, study the distribution of the Jordan block sizes, reporting exponentially many blocks of thermodynamically large size. 

{The rest of the article is organized as follows. In Sec.~\ref{sec:IM_approach} we recall the IM approach to dynamical systems. In Sec.~\ref{sec:YB_Transfer_matrix} we introduce a family of integrable transfer matrices that admit the IM as a unique joint eigenvector with eigenvalue one. In Sec.~\ref{sec:BA} we describe the structure of the Bethe ansatz equations (BAE) and show that the IM is realized as a particular Bethe vector with analytic Bethe roots. Finally, in Sec.~\ref{sec:free_fermionic_point} we specialize to the free-fermionic limit: we introduce an additional family of local Hamiltonians commuting with the transfer matrices, study the distribution of Jordan-block sizes, and identify a basis of quasi-localized particles. This basis, in turn, allows us to construct a quasi-local gapped parent Hamiltonian.}
\section{IM approach to Floquet dynamics}\label{sec:IM_approach}
\begin{figure}[t]
    \centering
    \includegraphics[width=0.9\linewidth]{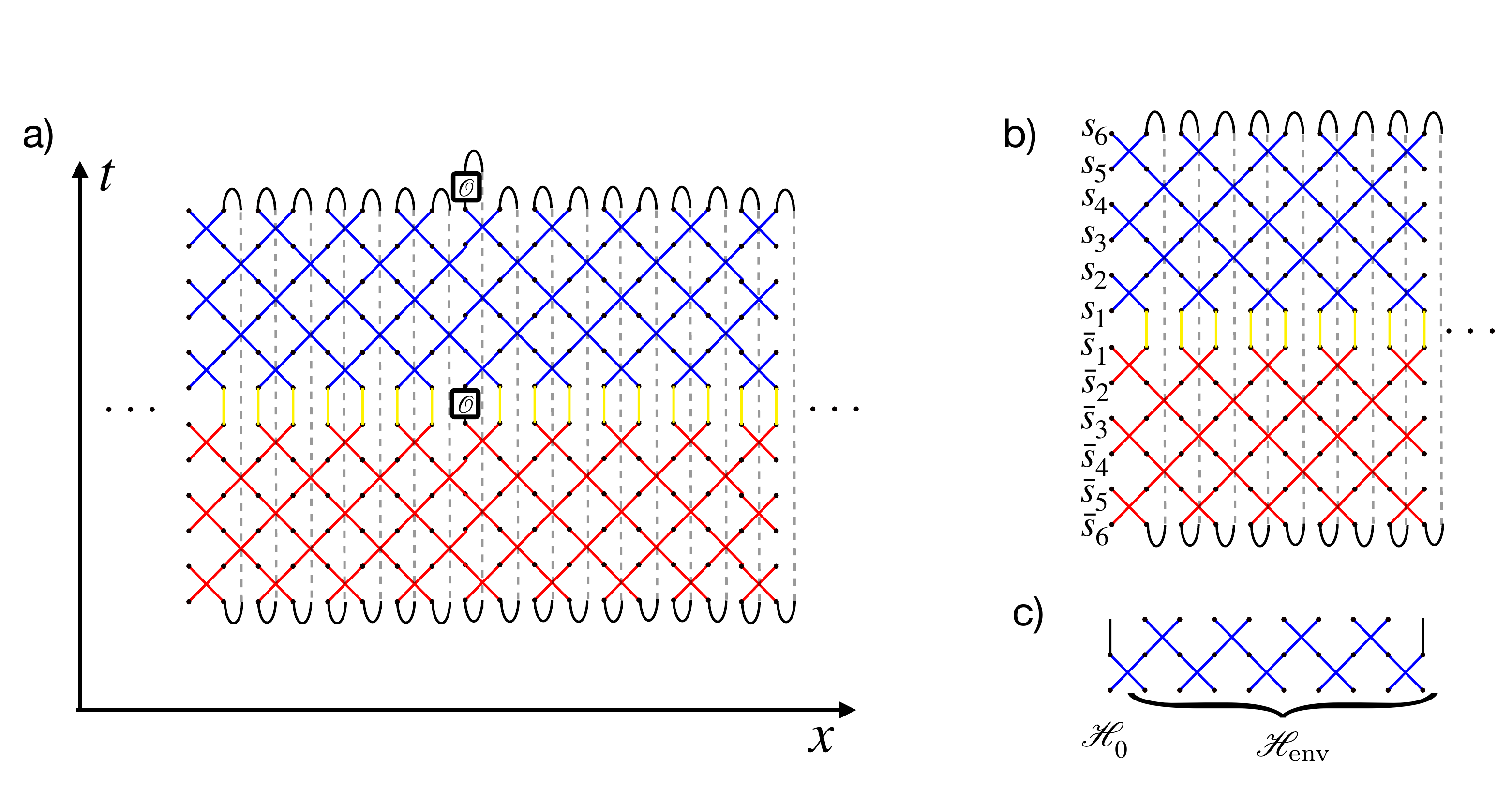}
    \caption{a) Pictorial diagram for a two-point function, the yellow lines correspond to a twisted initial density matrix $\rho_b$ from Eq.~\eqref{eq:rho_b}. b) The definition of the IM, with fixed spin trajectory and $N=3$. c) Operator $U^{\textrm{right}}$ acting on two Hilbert spaces $\mathscr{H}_0$ and $\mathscr{H}_{\textrm{env}}$ from Eq.~\eqref{eq:U_half}}
    \label{fig:IM_definition}
\end{figure}
\paragraph{Brick-wall Floquet dynamics, and IM}
Let us consider an infinite spin chain $\mathscr{H}= \cdots\mathbb{V}_{-2}\otimes\mathbb{V}_{-1}\otimes\mathbb{V}_0\otimes \mathbb{V}_1\otimes \mathbb{V}_2\dots$, where each space $\mathbb{V}_s$ is isomorphic to $\mathbb{C}^2$. The chain is equipped with two-qubit gates $U_{i,i+1}$ - unitary operators acting non-trivially only on the $i$-th and $(i+1)$-th spins. We define two evolution operators 
$U_{\text{odd}}=\prod\limits_{i=-\infty}^{\infty}U_{2i-1,2i}$, $U_{\text{even}}=\prod\limits_{i=-\infty}^{\infty}U_{2i,2i+1}$, and their product $U=U_{\text{even}}U_{\text{odd}}$. 

{In this article, we consider a concrete two-qubit unitary gate $U_{i,i+1}$ given by the well-known $\mathrm{XXZ}$ $R$-matrix:
\begin{equation}\label{eq:R-matrix}
    R_{i,i+1}(u)=\begin{pmatrix}
        1 & 0&0 & 0\\
        0& \frac{\sinh(u)}{\sinh(u+\eta)}& \frac{\sinh(\eta)}{\sinh(u+\eta)}&0\\
        0& \frac{\sinh(\eta)}{\sinh(u+\eta)}& \frac{\sinh(u)}{\sinh(u+\eta)}&0\\
        0 & 0&0 & 1
    \end{pmatrix}_{i,i+1},
\end{equation}
where, in order to ensure unitarity condition, we treat $u$ as a real parameter, while $\eta$ is purely imaginary.}
This operator sets a non-trivial dynamics that might be thought of as a Trotterization \cite{vanicat2018integrable} of the XXZ dynamics \cite{bulchandani2018bethe,ljubotina2019ballistic}. In this article, we will study the approach of the influence matrix (IM) and discover the structures of integrability in this case. 

{
The dynamics can be represented pictorially as in
Fig.~\ref{fig:IM_definition}(a), where, for concreteness, we consider
the dynamical two-point function
$\langle \mathcal{O}(N)\mathcal{O}(0)\rangle$.
Throughout this work, $N$ denotes the number of time steps.
We restrict ourselves to initial bulk density matrices of
product-state form (depicted in yellow):
\begin{equation}\label{eq:rho_b}
\rho_b=
\frac{q^{\sigma_z}}{q+q^{-1}}
\otimes
\frac{q^{\sigma_z}}{q+q^{-1}}
\otimes\cdots,
\qquad q>0.
\end{equation}
The twist parameter $q$ controls the average magnetization of
the initial state: the limits $q\to\infty$ and $q\to0$ correspond
to fully polarized states with all spins up and all spins down,
respectively, while $q=1$ yields the maximally mixed state.
Owing to the $U(1)$ symmetry of the two-qubit gates $U_{i,i+1}$,
this homogeneous product state is stationary under the bulk
time evolution. This family of initial states also preserves
the temporal integrability introduced below.
}

The IM approach is based on partitioning the system into a
small subsystem and an environment. {Here, we consider a semi-infinite chain, with the spin at its
boundary, site $0$, serving as the subsystem and the sites $i>0$
constituting the environment, as shown in
Fig.~\ref{fig:IM_definition}(c).}
We introduce the half-chain brickwork operators
\begin{equation}\label{eq:U_half}
U^{\textrm{right}}_{\text{odd}}
=\prod_{i=1}^{\infty}U_{2i-1,2i},
\qquad
U^{\textrm{right}}_{\text{even}}
=\prod_{i=0}^{\infty}U_{2i,2i+1},
\qquad
U^{\textrm{right}}
=U^{\textrm{right}}_{\text{odd}}\,
 U^{\textrm{right}}_{\text{even}}.
\end{equation}
The operator $U^{\textrm{right}}$ acts on
$\mathscr{H}_0\otimes\mathscr{H}_{\mathrm{env}}$,
the tensor-product Hilbert space of the subsystem and the
environment.

{
To define the IM, we fix trajectories of the boundary spin
on the forward and backward contours,
$\boldsymbol{s}=\{s_1,s_2,\dots,s_{2N}\}$ and
$\bar{\boldsymbol{s}}=\{\bar{s}_1,\bar{s}_2,\dots,\bar{s}_{2N}\}$,
respectively. Odd and even indices label incoming and outgoing
legs, respectively. Each time step $i$ therefore involves four
spin variables,
$\{s_{2i-1},s_{2i},\bar{s}_{2i-1},\bar{s}_{2i}\}$,
as depicted in Fig.~\ref{fig:IM_definition}(b), with each variable
taking values $\pm1$.
Tracing out the environmental degrees of freedom yields the
IM functional of the boundary trajectories
$(\boldsymbol{s},\bar{\boldsymbol{s}})$, defined as
}
\begin{multline}\label{IMDef}
\mathcal{I}_{\boldsymbol{s}}^{\bar{\boldsymbol{s}}}
=
\operatorname{tr}_{\mathrm{env}}\Big(
\langle \bar{s}_{2N}|(U^{\textrm{right}})^{-1} |\bar{s}_{2N-1}\rangle
\cdots
\langle \bar{s}_{2}|(U^{\textrm{right}})^{-1}|\bar{s}_1\rangle
\\
\times\,
\rho_b\,
\langle s_1|U^{\textrm{right}}|s_2\rangle
\langle s_3|U^{\textrm{right}}|s_4\rangle
\cdots
\langle s_{2N-1}|U^{\textrm{right}}|s_{2N}\rangle
\Big).
\end{multline}
Here $\operatorname{tr}_{\mathrm{env}}$ denotes the trace over all environment degrees of freedom, while the bra-ket matrix elements are taken with respect to the subsystem spin.

The IM contains information about the correlation functions of arbitrary operators $\mathcal{O}$ acting locally on the subsystem spin. For example, the two point correlation function $\langle\mathcal{O}(N)\mathcal{O}(0)\rangle$ can be computed by sandwiching it between the left and right IMs as can be {seen} in Fig.~\ref{fig:IM_definition} a):
{\begin{gather}\label{CorrIM}
\langle\mathcal{O}(N)\mathcal{O}(0)\rangle=\sum\limits_{s_i\bar{s}_i} (\mathcal{I}_l)^{\bar{s}_1,\dots,\bar{s}_{2N}}_{s_1,\dots,s_{2N}} (\mathcal{I}_r)^{\bar{s}_{0},\dots,\bar{s}_{2N-1}}_{s_0,\dots,s_{2N-1}}\mathcal{O}_{s_0,\bar{s}_0} \mathcal{O}_{s_{2N},\bar{s}_{2N}}.
\end{gather}}
here $\mathcal{I}_r$ is given by the same formula as in \eqref{IMDef}, while $\mathcal{I}_l$ is constructed in the same manner, for the evolution operator corresponding to the semi-infinite bath spreading to the left, {and the operator evolution is governed by the equation $\mathcal{O}(t+1)=U^{-1}\mathcal{O}(t) U$}.

Let us vectorize the IM, treating it as a vector, {Eq.~\eqref{CorrIM} might be rewritten in a bra-ket notation:}
\begin{equation}\label{eq:one-point correlator}
{\langle \mathcal{O}(N)\mathcal{O}(0)\rangle
= \langle \mathcal{I}_l\,\mathcal{O}^{\perp}_{N}\,|\,\mathcal{O}^{\perp}_0\,\mathcal{I}_r\rangle.}
\end{equation}

{Note that the space-dual counterpart of the operator $\mathcal{O}$ is a projector, acting on temporal degrees of freedom, denoted by $\mathcal{O}^{\perp}_k$, reducing the number of indices of the IM and is defined as follows:}
\begin{gather}
{\mathcal{O}^{\perp}_k|s_k,\bar{s}_k\rangle=\mathcal{O}_{s_{k},\bar{s}_{k}}.}
\end{gather}
Note that left and right IM may have a different initial state $\rho_b$ allowing us to study the quench dynamics. Another possible application - is to consider left and right IMs corresponding to different integrable systems (for example, systems with different anisotropy $\Delta$). In this case, the correlator \eqref{eq:one-point correlator} corresponds to a non-integrable dynamics, but still can be calculated from the integrable data.

As we argue in this text, the temporal approach allows one to treat the IM as a special Bethe vector, effectively translating dynamical correlations into static ones with a particular Bethe vector.
\subsection{Causality and Physicality}\label{sec:Caus_Phys}
Before we proceed to define integrability, let us mention a few important properties of the IM. First, using the Choi operator-state duality, one may consider the $\mathcal{I}_{\boldsymbol{s}}^{\bar{\boldsymbol{s}}}$ as a density matrix of a certain state:
\begin{eqnarray}\label{positivity}
\left(\rho^{\mathcal{I}}\right)_{\boldsymbol{s}}^{\bar{\boldsymbol{s}}}=2^{-2N}\mathcal{I}_{\boldsymbol{s}}^{\bar{\boldsymbol{s}}}.
\end{eqnarray}
This means that the IM can be considered an operator, which is positive, with the trace equal to one. Another property is a reduction:
\begin{eqnarray}\label{reduction}
\sum\limits_{s_n=\bar{s}_n}\mathcal{I}_{s_1,\dots s_n}^{\bar{s}_1,\dots \bar{s}_n}=\delta_{s_{n-1},\bar{s}_{n-1}}\mathcal{I}_{s_1,\dots s_{n-2}}^{\bar{s}_1,\dots \bar{s}_{n-2}}.
\end{eqnarray}
This second property reflects the causality principle: measurements taken at later times cannot affect the dynamics at earlier times. This observation is best seen from the following simplified graphical representation of the IM.
\paragraph{Causality light cone}
In order to consider dynamics up to $N$ time steps, it suffices to consider {first} $N$ sites of the bath only, as sites beyond this range do not influence the dynamics. The latter proof could be readily illustrated pictorially.
\begin{figure}[H]
    \centering
    \includegraphics[scale=0.18]{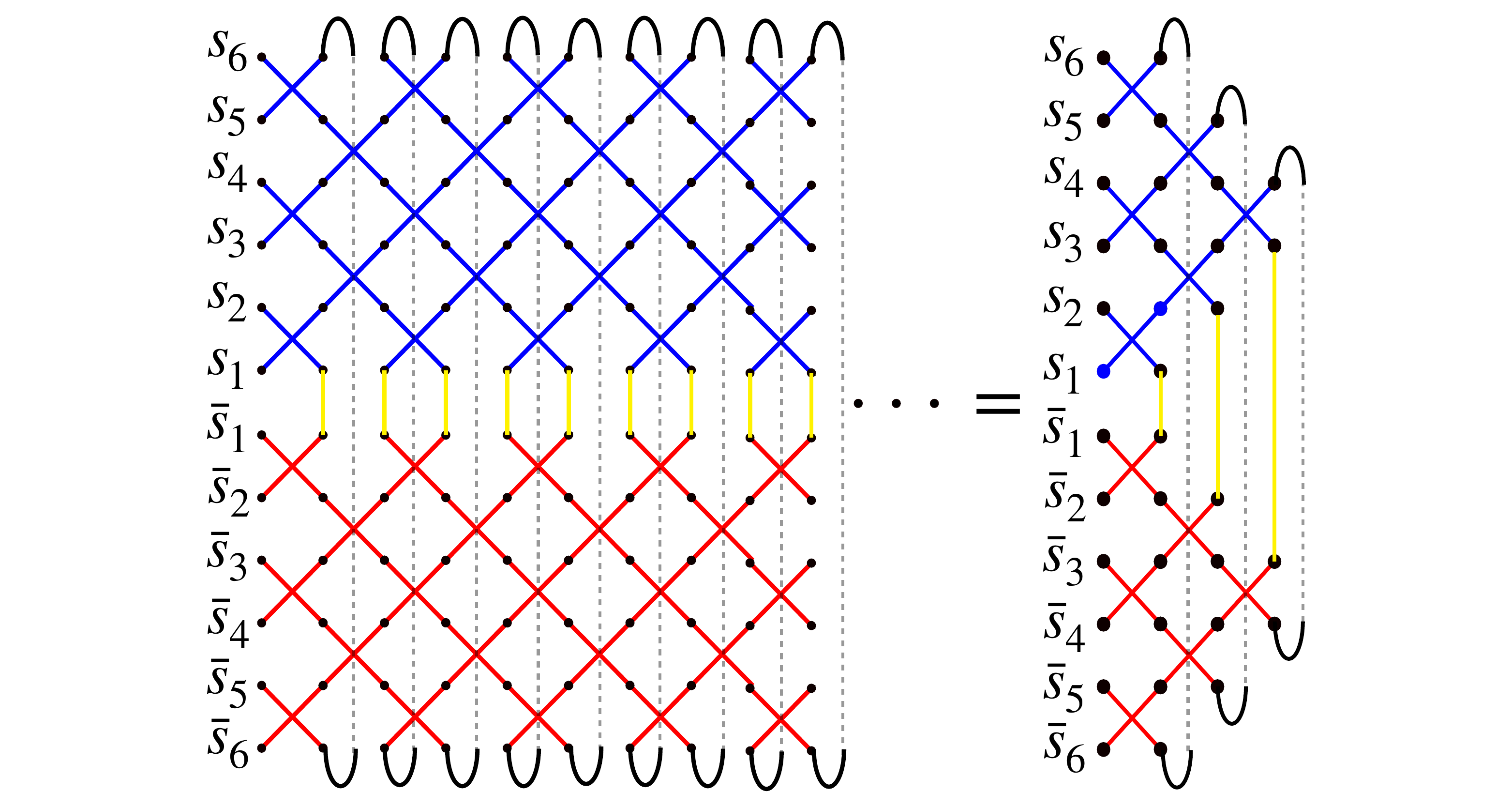}
    \caption{Pictorial representation of the causality.}
    \label{fig:IM}
\end{figure}
where we used the identities $U_{i,i+1} U^{-1}_{i,i+1}=\mathrm{Id}$ and
$U_{i,i+1}q^{\sigma^{z}_i+\sigma^{z}_{i+1}} U^{-1}_{i,i+1}=q^{\sigma^{z}_i+\sigma^{z}_{i+1}}$, namely
\begin{subequations}\label{eq:unitarity_pictorial}
\begin{align}
\label{eq:unitarity_pictorial_a}
&\raisebox{-0.5\height}{\includegraphics[scale=0.3]{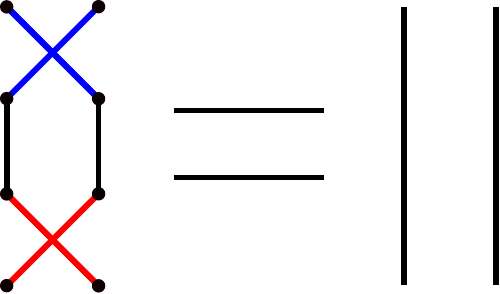}}
\\[0.6em]
\label{eq:unitarity_pictorial_b}
&\raisebox{-0.5\height}{\includegraphics[scale=0.3]{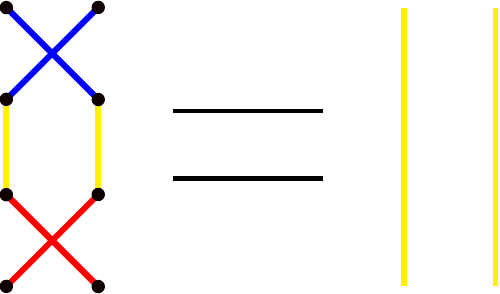}}
\end{align}
\end{subequations}
\paragraph{Temporal transfer matrix}
It is always possible to introduce a temporal transfer matrix $\mathcal{T}$, such that it stabilizes the IM: $\mathcal{T}|\mathcal{I}\rangle=|\mathcal{I}\rangle$. Pictorially, one can define it as follows:
\begin{figure}[H]
    \centering
    \includegraphics[scale=0.12]{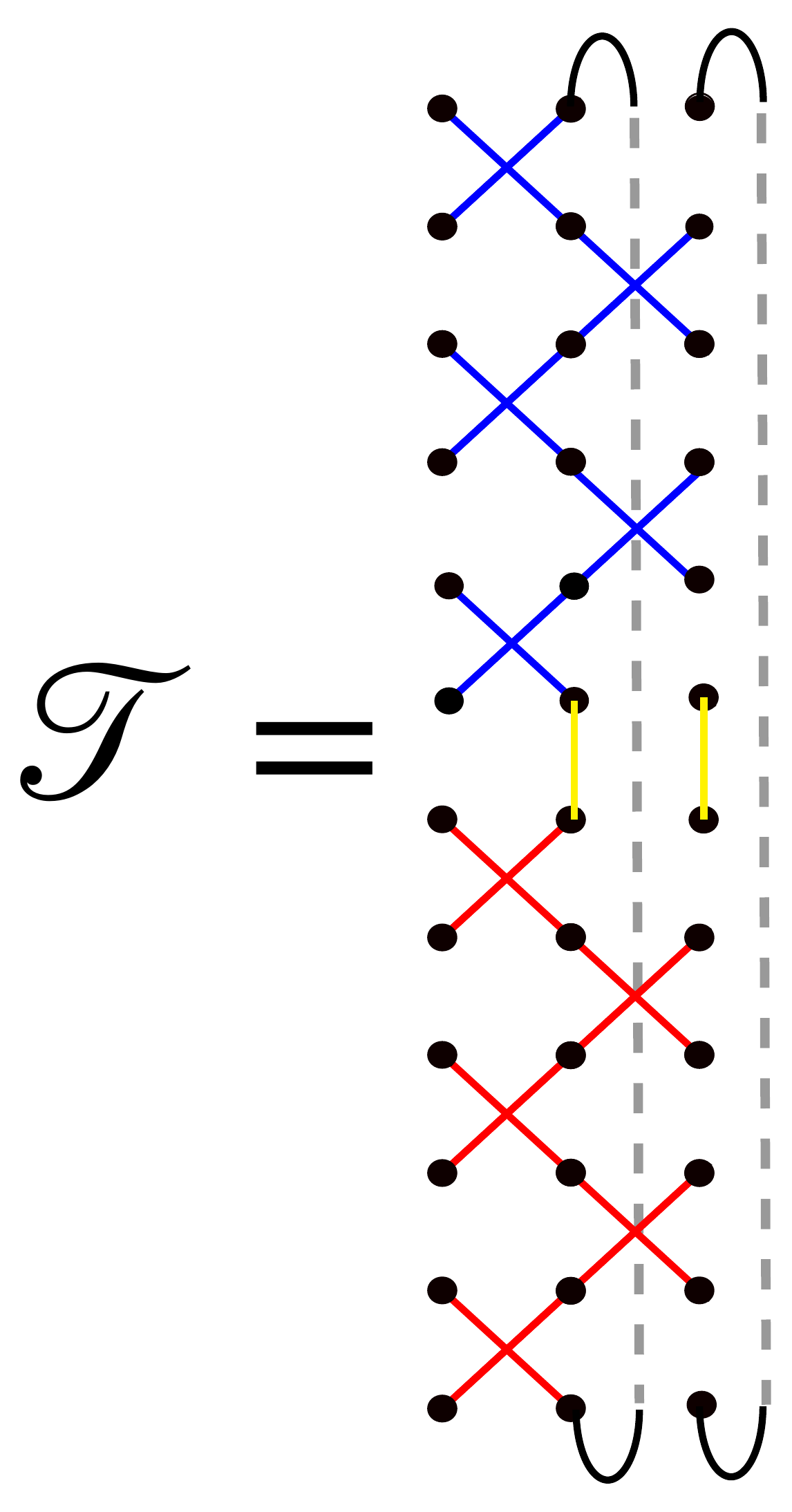}
    \caption{The graphical definition of the transfer matrix.}
    \label{fig:Transfer_matrix}
\end{figure}
Properties \eqref{positivity},\eqref{reduction} have their counterparts at the level of the transfer matrix. Namely, one can ensure that the temporal transfer matrix maps influence matrices to influence matrices. {It is also clear that the unitarity condition, provided by Eqs~(\ref{eq:unitarity_pictorial_a}--\ref{eq:unitarity_pictorial_b}) guarantees
$\operatorname{tr}\!\left(\mathcal{T}^n\right)=1$ for any $n\ge 1$.
This is a strong constraint, since it allows one to determine the characteristic polynomial explicitly:
\begin{equation}
\det\!\left(x-\mathcal{T}\right)=(x-1)x^{2^{4N}-1}.
\end{equation}
In particular, $\mathcal{T}$ has a single eigenvalue {that} equals $1$ - corresponding to the IM, and the remaining $2^{4N}-1$ eigenvalues are $0$.
Consequently, $\mathcal{T}$ is generically non-diagonalizable and must contain nontrivial Jordan blocks associated with the degenerate eigenvalue $0$. As we will see in the next section, in the integrable case, this transfer matrix can be generalized to a commuting family of transfer matrices, such that the IM is their joint eigenvector with eigenvalue one.}

\section{Yang-Baxter relation and auxiliary transfer matrix.}\label{sec:YB_Transfer_matrix}
In this article, we consider a concrete two-qubit unitary gate $U_{i,i+1}$ given by the well-known $\mathrm{XXZ}$ $R$-matrix.

In this section, we construct an integrable transfer matrix such that the IM is its eigenvector with the largest eigenvalue, equal to $1$. Before doing so, let us introduce some notation. First, it is convenient to assign a spectral parameter to each line. At the intersection of two lines carrying spectral parameters $u$ and $v$, we place an $R$-matrix that depends on their difference, $R(u-v)$. These conventions are illustrated in the figures below:
\begin{eqnarray}
\psfrag{u}{$u$}
\psfrag{v}{$v$}
\psfrag{=}{$=$}
\psfrag{R}{$R(u-v)$}
    \includegraphics[scale=0.4]{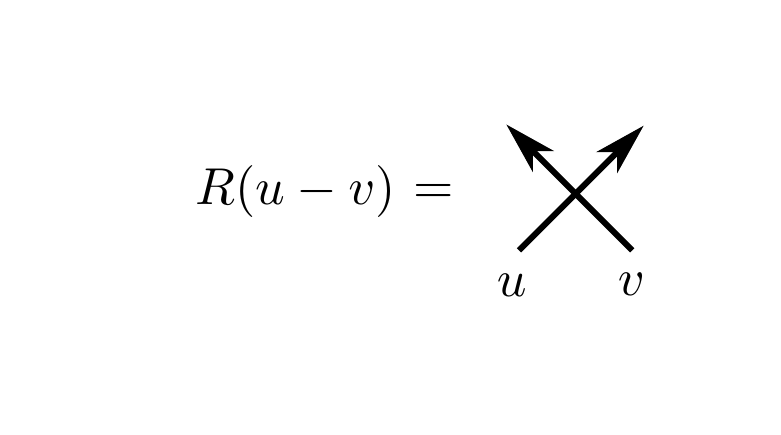}
    \label{fig:Rmat}
\end{eqnarray}
\begin{equation}
\psfrag{u}{$u$}
\psfrag{0}{$0$}
    \includegraphics[scale=0.4]{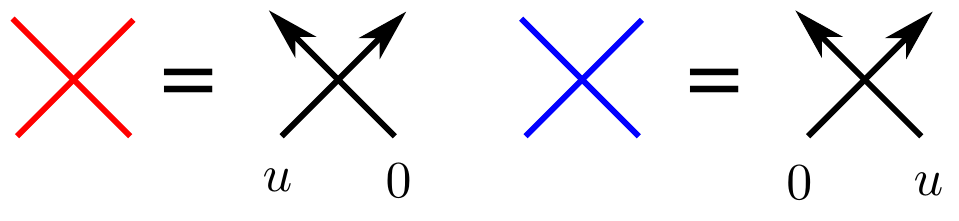}
\end{equation}
The unitarity condition from Eq.~\eqref{eq:unitarity_pictorial_a} simply rewrites as follows:
\begin{figure}[H]
\psfrag{u}{$u$}
\psfrag{v}{$v$}
    \centering
    \includegraphics[scale=0.2]{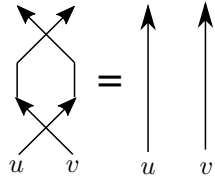}
    \caption{Unitarity condition in an $R-$matrix notations.}
    \label{fig:unitarity2}
\end{figure}

The nice property of the $R-$matrix is, of course, the Yang-Baxter relation:
\begin{equation} \label{YB}
R_{1,2}(v_1-v_2)R_{1,3}(v_1-v_3)R_{2,3}(v_2-v_3)=R_{2,3}(v_2-v_3)R_{1,3}(v_1-v_3)R_{1,2}(v_1-v_2),
\end{equation}
or after the transposition with respect to site $3$:
\begin{equation}\label{YBt}
R_{1,2}(v_1-v_2)R^{t_3}_{2,3}(v_2-v_3)R_{1,3}^{t_3}(v_1-v_3)=R_{1,3}^{t_3}(v_1-v_3)R_{2,3}^{t_3}(v_2-v_3)R_{1,2}(v_1-v_2).
\end{equation}
The latter can be depicted pictorially as follows:
\begin{eqnarray}
\psfrag{0}{$u_3$ }
\psfrag{u}{$u_1$}
\psfrag{v}{$u_2$}
\includegraphics[scale=0.4]{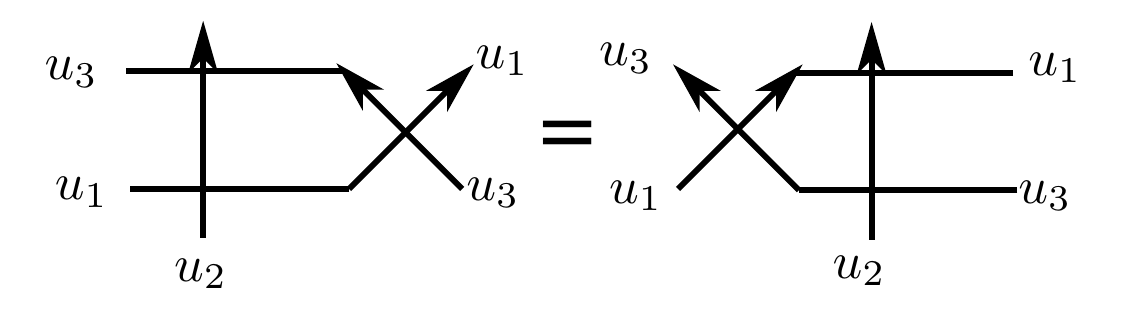}
\end{eqnarray}

It is convenient to equip our pictorial definition of the IM from Fig.~\ref{fig:IM} with arrows and spectral parameters, as depicted below:
\begin{equation}
\includegraphics[scale=0.15]{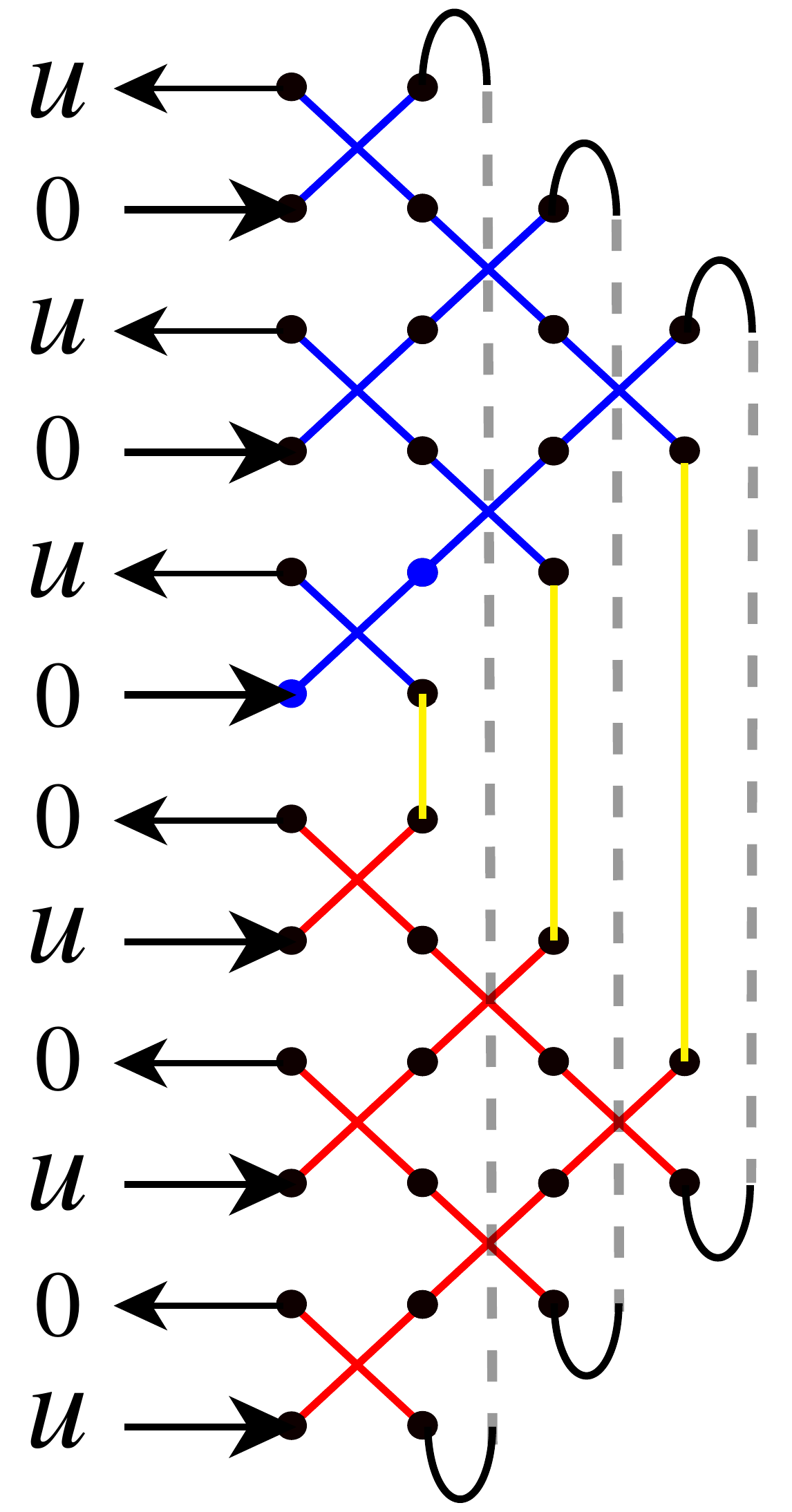}.
\end{equation}
It is natural to consider the following transfer matrix:
\begin{multline}\label{eq:Transfer_matrix}
    T(v)=\operatorname{tr}_0 \Big(R_{4N,0}(u-v)R^{t_{4N-1}}_{0,4N-1}(-v)\dots R_{2N+2,0}(u-v)R^{t_{2N+1}}_{0,2N+1}(v)\times \\ \times \rho_0(q) R_{2N,0}(-v)R^{t_{2N-1}}_{0,2N-1}(v-u)\dots R_{2,0}(-v)R^{t_{1}}_{0,1}(v-u)\Big),
\end{multline}
here $\rho_0(q)=\frac{q^{\sigma_0^z}}{q+q^{-1}}$. This transfer matrix can also be defined pictorially:
\begin{eqnarray}
\includegraphics[scale=0.15]{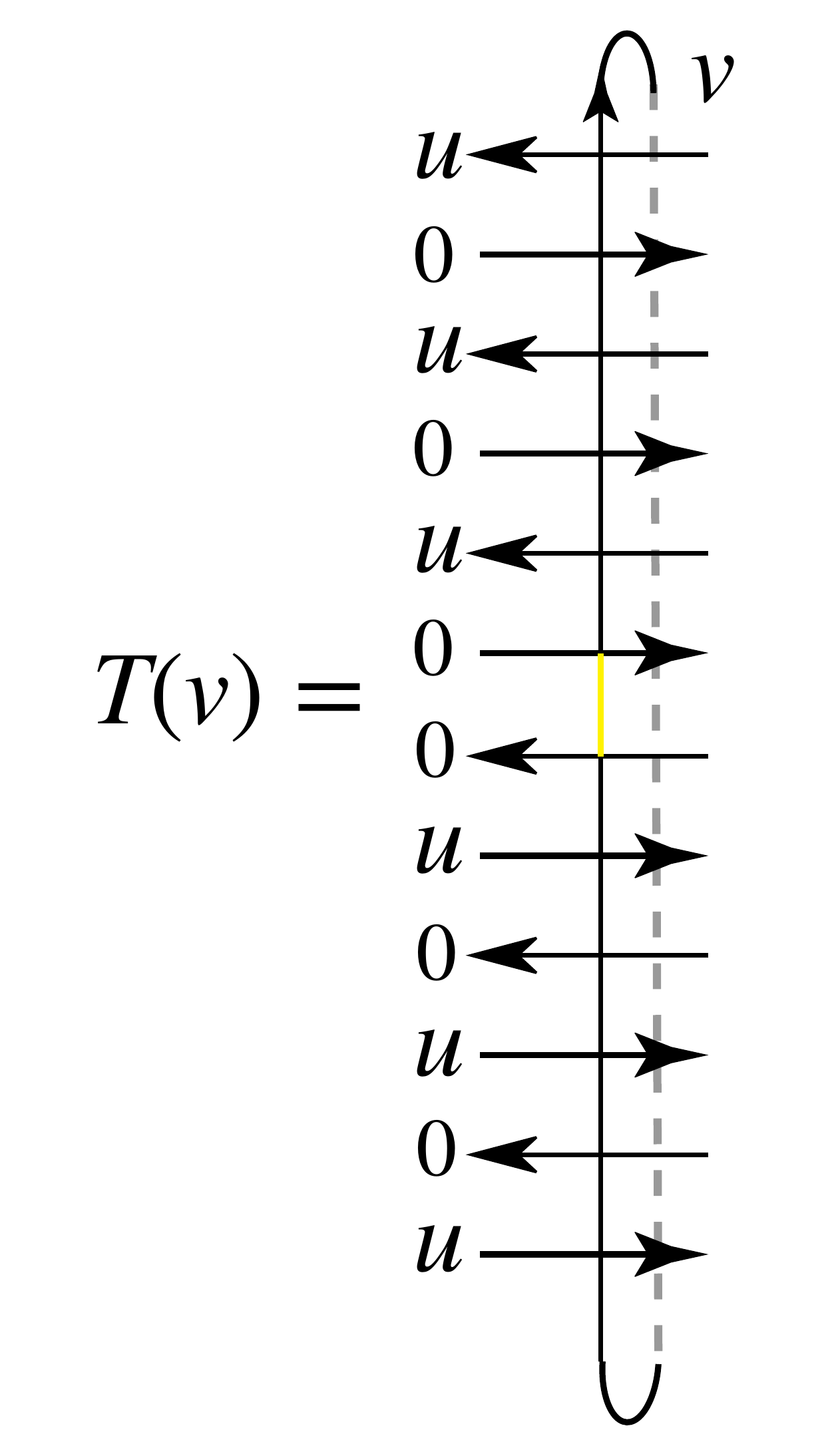}
\end{eqnarray}
Note that the transfer matrix is a completely auxiliary object, at first glance, it has nothing to do with the IM. We see, however, that the IM is an eigenvector of $T(v)$ with eigenvalue one.\\
The proof is very simple and could be done pictorially:
\begin{figure}[H]
    \centering
    \includegraphics[scale=0.15]{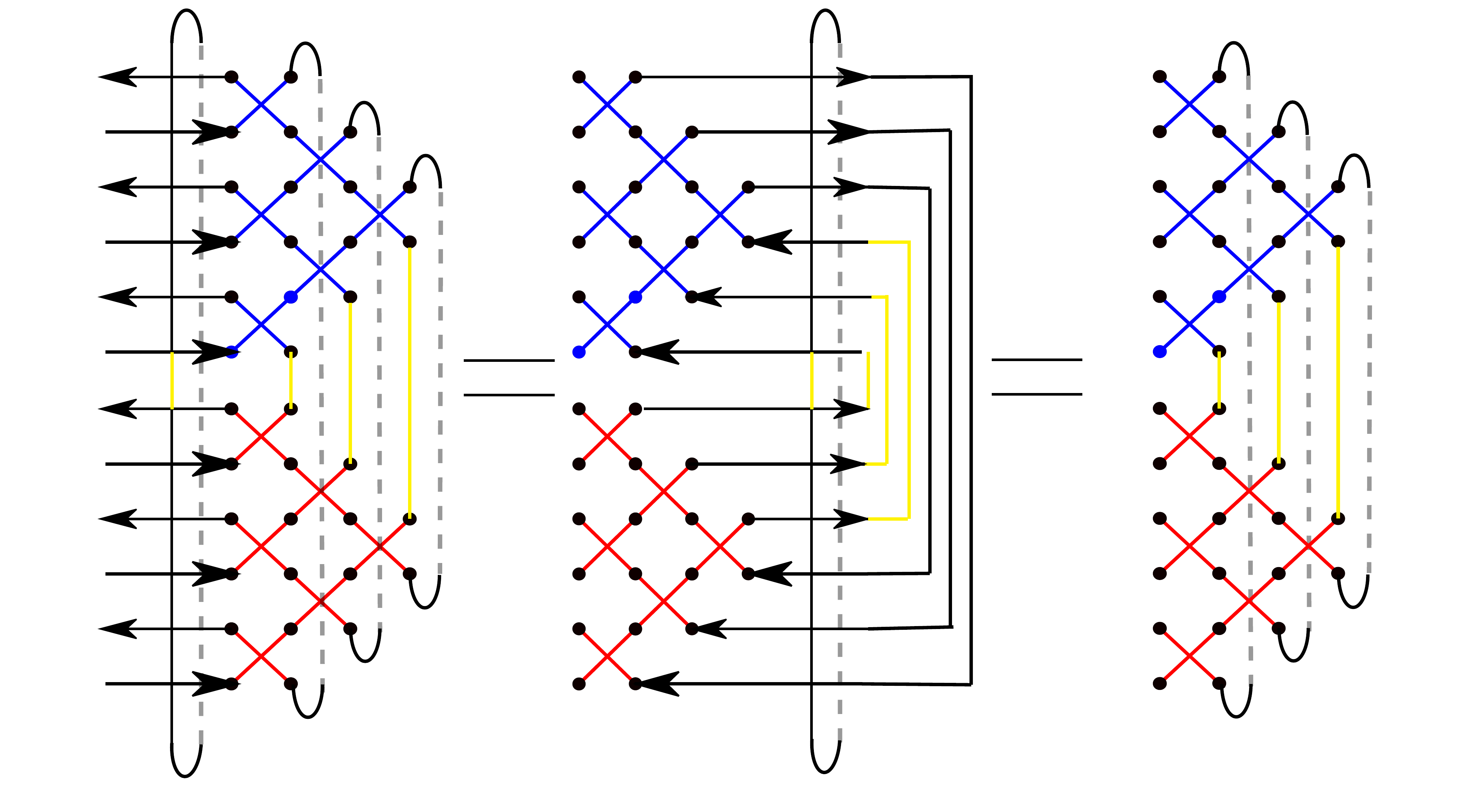}
    \caption{Definition of $T$ and pictorial proof of the equation $T|\mathcal{I}\rangle=|\mathcal{I}\rangle$.}
    \label{fig:Tmatrix}
\end{figure}
In the first identity, we used Yang-Baxter \eqref{YBt}, while in the second identity, we used the unitarity condition \eqref{fig:unitarity2}.
\\
Our goal for this article is to develop the Bethe ansatz techniques for the influence matrix.\\
The nice property is that \eqref{YB},\eqref{YBt} implies the commutation of
transfer matrices of different arguments:
\begin{equation}
[T(v_1),T(v_2)]=0.
\end{equation}
The commutativity again follows from the Yang-Baxter relations \eqref{YB},\eqref{YBt}. {The temporal transfer matrix $\mathcal{T}$  itself also belongs to this commuting family and can be written exactly as:}
\begin{equation}\label{eq:Tau}
\mathcal{T}=T(0)T(u).
\end{equation}
{This equation follows directly from the definition of transfer matrices, and careful specialization to special points $v=\{0,u\}$, at which some of the $R$-matrices are equal to permutation operators.}
As we have already seen $\mathcal{T}$ matrix is generically non-diagonalizable, and as we argue later, although the generalized transfer matrices $T(v)$ resolve some of the degeneracies, they also remain non-diagonalizable. 
\paragraph{Crossing relation, equivalent transfer matrix, degenerate modules.}
In order to get rid of unpleasant transpositions, and bring the transfer matrix to a more convenient form, we use the crossing relation:
\begin{equation}
R_{k,0}^{t_k}(-v)=\frac{\sinh(v)}{\sinh(v-\eta)}\sigma^y_k R_{0,k}(v-\eta)\sigma^y_k.
\end{equation}
After the additional transposition over the auxiliary site $0$ and the conjugation by an operator $\prod\limits_{i=1}^{2N}\sigma_{2i-1}^y$, the transfer matrix in $\eqref{eq:Transfer_matrix}$ is similar to the following transfer matrix $\tilde{T}(v)=\prod\limits_{i=1}^{2N}\sigma_{2i-1}^yT(v)\prod\limits_{i=1}^{2N}\sigma_{2i-1}^y$. This matrix is provided explicitly by:
\begin{multline}\label{TildeTM}
    \tilde{T}(v)=\left(\frac{\sinh(v)\sinh(v-u)}{\sinh(v-\eta)\sinh(v-u-\eta)}\right)^N \mathrm{tr}_0 \Big(R_{0,1}(v-u)R_{0,2}(v-\eta)\dots R_{0,2N-1}(v-u)\times \\\times R_{0,2N}(v-\eta) \rho_0 (q^{-1}) R_{0,2N+1}(v)R_{0,2N+2}(v-u-\eta)\dots R_{0,4N-1}(v)R_{0,4N}(v-u-\eta)\Big).
\end{multline}

{In order to gain some intuition about the structure of this transfer matrix, we consider the product of two $R$-matrices with spectral parameters shifted by $\pm\eta$. We will observe that this product has invariant symmetric/antisymmetric subspaces, but it cannot be decomposed into a direct sum of the two:
\begin{align}\label{eq:proj_1}
& P^{a} R_{0,1}(v+\eta)\,R_{0,2}(v)\,P^{s}=0 \, ,\\
& P^{s} R_{0,1}(v-\eta)\,R_{0,2}(v)\,P^{a}=0 \, , \label{eq:proj_2}\\
& P^{s} R_{0,1}(v+\eta)\,R_{0,2}(v)\,P^{a}\neq 0 \, , \label{eq:proj_3}\\
& P^{a} R_{0,1}(v-\eta)\,R_{0,2}(v)\,P^{s}\neq 0 \, , \label{eq:proj_4}
\end{align}
where $P^{s/a}$ denote the projectors onto the symmetric/antisymmetric subspaces.}

{The first two identities follow directly from the YBE,
\begin{equation}\label{eq:resonance_identity}
R_{2,1}(v-u)\,R_{0,2}(v)\,R_{0,1}(u)=R_{0,1}(u)\,R_{0,2}(v)\,R_{2,1}(v-u),
\end{equation}
by taking the limit $u\to v\pm\eta$. Indeed, the leading term of $R_{2,1}(\mp\eta)$ is a projector onto the symmetric/antisymmetric subspace. Therefore, the product $R_{0,1}(v-\eta)\,R_{0,2}(v)$ has an invariant symmetric subspace, while the product in the opposite order, $R_{0,1}(v+\eta)\,R_{0,2}(v)$, has an invariant singlet (antisymmetric) subspace.}

{The most direct way to obtain the last two relations is to compute matrix elements of the two-site Lax operators,
\begin{equation}\label{eq:two-site_lax}
L^{\pm}_{\alpha,\beta}(v)={}_0\langle \alpha|\,R_{0,1}(v\pm\eta)\,R_{0,2}(v)\,|\beta\rangle_0 \, .
\end{equation}
This calculation is carried out in Appendix~\ref{app:Tranfser matrix at degenerate points}, where we establish Eqs.~(\ref{eq:proj_3}--\ref{eq:proj_4}). Moreover, we observe that the standard Bethe ansatz is not applicable at the resonance point, because the operators $B^{\pm}(v)=L^{\pm}_{0,1}(v)$ do not generate the full space from the highest-weight vector. We will overcome this difficulty by introducing a small perturbation $\epsilon$ of the spectral parameters, which resolves the degeneracy and allows a Bethe-ansatz computation. At the end, of course, the limit $\epsilon\to 0$ must be taken carefully. We will also briefly show that degeneracies occur already for $N=2$, leading to non-diagonalizability and Jordan-block structure of the corresponding transfer matrices.
}
\section{Bethe ansatz}\label{sec:BA}
As was announced in previous section, and in order to use the standard Bethe ansatz technique, let us resolve the degeneracy by slightly shifting the spectral parameters. 
\begin{multline}\label{Tmat}
    \tilde{T}_{\epsilon}(v)=\left(\frac{\sinh(v)\sinh(v-u)}{\sinh(v-\eta)\sinh(v-u-\eta)}\right)^N \mathrm{tr}_0 \Big(R_{0,1}(v-u)R_{0,2}(v-\eta-\epsilon)\dots R_{0,2N-1}(v-u)\times \\\times R_{0,2N}(v-\eta-\epsilon) \rho_0(q^{-1}) R_{0,2N+1}(v)R_{0,2N+2}(v-u-\eta-\epsilon)\dots R_{0,4N-1}(v)R_{0,4N}(v-u-\eta-\epsilon)\Big).
\end{multline}
It is well known that in this case one should express the eigenvectors in terms of the matrix elements of the monodromy matrix:
\begin{multline} \label{MonodromyMatrix}
    L_{\epsilon}(v)=  \Big(R_{0,1}(v-u)R_{0,2}(v-\eta-\epsilon)\dots R_{0,2N-1}(v-u)R_{0,2N}(v-\eta-\epsilon)\times \\ \times \rho_0(q^{-1}) R_{0,2N+1}(v)R_{0,2N+2}(v-u-\eta-\epsilon)\dots R_{0,4N-1}(v)R_{0,4N}(v-u-\eta-\epsilon)\Big)
\end{multline}
\begin{equation} \label{ABCD}
   L_{\epsilon}(v-\frac{\eta}{2})=\begin{pmatrix} A_{\epsilon}(v) && B_{\epsilon}(v) \\
   C_{\epsilon}(v) && D_{\epsilon}(v)
   \end{pmatrix}_0,
\end{equation}
here, matrices $A,B,C,D$ act in the Hilbert space of the spin chain, and $L_{\epsilon}(v)$ is considered as a $2 \times 2$ matrix in the auxiliary zero space.
\\
The eigenvectors are given by a multiple product of $B_{\epsilon}(x)$ operators acting on the pseudo vacuum $ |\uparrow\dots\uparrow\rangle$
\begin{equation}\label{eq:Bvec}
|\mathbb{B}_{\epsilon}(x_1,\dots x_n)\rangle=B_{\epsilon}(x_1)\dots B_{\epsilon}(x_n)|\uparrow\dots\uparrow\rangle.
\end{equation}
The conditions for $B_{\epsilon}(x_1,\dots x_n)$ to be an eigenvector form the system of non-linear Bethe ansatz equations (BAE). {While these equations are analyzed in detail in Appendix~\ref{app:BAE}, here we outline the main ideas and refer the interested reader there for a complete derivation. Starting from the Bethe ansatz equations~\eqref{eq:BAE}, one can rewrite them in the form}
{\begin{multline} \label{eq:BAEM_main}
\Bigg(\frac{\sinh(x_i +\epsilon)}{\sinh(x_i)}\frac{\sinh(x_i - u +\epsilon)}{\sinh(x_i - u)} \frac{\sinh(x_i -\eta)}{\sinh(x_i +\eta)} \frac{\sinh(x_i - u -\eta)}{\sinh(x_i - u +\eta)} \Bigg)^{N} =
\\= -q^{-2} \prod\limits_{j\neq i} \frac{\sinh(x_i - x_j + \eta)}{\sinh(x_i-x_j-\eta)},
\end{multline}}
{where the $\epsilon$-dependent factors that may lead to singular behavior in the limit $\epsilon \to 0$ are made explicit.}

{For Bethe roots that stay away from the special points $x=0,u$, the $\epsilon$-dependent factors in \eqref{eq:BAEM_main} can be neglected, and the equations reduce to those of a spin-one chain on $2N$ sites. This reproduces the sector containing only spin-one degrees of freedom, discussed at the end of the previous section.}

{However, the limit is singular in the vicinity of the resonance points $x=0$ and $x=u$, where the $\epsilon$-dependent factors produce nontrivial contributions. This leads to additional solutions, which we parametrize by the scaling ansatz $x^{(u)}_k = u + \epsilon \chi_k + o(\epsilon)$ and $x^{(0)}_k = \epsilon \xi_k + o(\epsilon)$.}

{Substituting this ansatz into \eqref{eq:BAEM_main}, and after fixing the numbers of generic roots $n_1$ and resonant roots $n_u$ and $n_0$, one obtains the system of three coupled equations: (i) the nontrivial Eq.~\eqref{eq:BAE 1} describing generic (spin-one) roots; and (ii) two Eqs.~\eqref{eq:BAE 2}--\eqref{eq:BAE 3} describing the resonant roots. The latter correspond to ``light'' spin-zero excitations, which can be solved perturbatively in $\epsilon$, and are responsible for the additional degeneracies appearing in the $\epsilon \to 0$ limit. We also emphasize that this limit must be taken with care: in naive expressions, some Bethe roots coincide, and higher-order corrections become essential. These corrections encode the Jordan-block structure. In particular, resolving the $k$-th generalized eigenvector within a Jordan cell requires expanding up to order $\epsilon^k$.}

{The influence matrix (IM) corresponds to a special, non-degenerate Bethe state built entirely from resonant Bethe roots. This can be seen from the defining property that it is the unique eigenvector with eigenvalue $1$. To verify this, let us examine the transfer-matrix eigenvalue in terms of the Bethe roots $x_i$:}
\begin{multline}\label{eq:TE_eigval}
\tau(v)=\frac{1}{q+q^{-1}}
\left(\frac{\sinh(v)\sinh(v-u)}{\sinh(v-\eta)\sinh(v-u-\eta)}\right)^N
\Bigg(
qr(v)\prod\limits_{i=1}^{2N}\frac{\sinh(x_i-v+\eta)}{\sinh(x_i-v)}
\\
+q^{-1}\prod\limits_{i=1}^{2N}\frac{\sinh(v-x_i+\eta)}{\sinh(v-x_i)}
\Bigg),
\end{multline}
{where $r(v)$ is the same function as on the left-hand side of Eq.~\eqref{eq:BAE}:}
\begin{equation}
r(v)=
\Bigg(\frac{\sinh(v+\epsilon)}{\sinh(v)}
\frac{\sinh(v-u+\epsilon)}{\sinh(v-u)}
\frac{\sinh(v-\eta)}{\sinh(v+\eta)}
\frac{\sinh(v-u-\eta)}{\sinh(v-u+\eta)}
\Bigg)^{N}.
\end{equation}

{The IM is the leading eigenvector and satisfies $\tau(v)\equiv 1+o(\epsilon)$ for all values of the spectral parameter $v$. This requirement is very restrictive and can only be met in the sector without spin-one particles, i.e., $n_u=n_0=N$. Indeed, consider the limit $v\to 0$. Equation~\eqref{eq:TE_eigval} contains a prefactor that vanishes as $(\sinh(v))^N$, while $r(v)$ contributes a compensating factor behaving as $(\sinh(v+\epsilon)/\sinh(v))^N\sim (v+\epsilon)^N/v^N$. Therefore $\tau(v)$ can remain finite (and equal to $1$) only if the products in Eq.~\eqref{eq:TE_eigval} supply an additional $v^{-N}$ singularity, which requires precisely $N$ Bethe roots approaching $0$ as $\epsilon\to 0$. Equivalently, we must have $N$ roots of the form $x_k^{(0)}=\epsilon\,\xi_k$. The same reasoning applied to the limit $v\to u$ enforces another set of $N$ roots approaching $u$, namely $x_k^{(u)}=u+\epsilon\,\xi_k$. A careful substitution of Eqs.~\eqref{on-shell_bethe_roots1}--\eqref{on-shell_bethe_roots2} then yields $\tau(v)=1$ in the limit $\epsilon\to 0$.} As a result, one has exactly:
\begin{eqnarray} \label{Bethe_Vector}
|\mathcal{I}\rangle=\frac{1}{(1+q^{2})^{2N}}\left(\frac{\epsilon}{\sinh(\eta)}\right)^{N^2}|\mathbb{B}_{\epsilon}(\vec{x}_{\epsilon})\rangle\ \Big|_{\epsilon\to 0},
\end{eqnarray}
where the Bethe roots are given explicitly by:\footnote{In principle, these formulas have to be corrected with higher powers of $\epsilon$, as we have checked numerically, these solutions provide the exact IM up to $4N=16$ sites. We conjecture these formulas to be exact with no further corrections needed.}
\begin{eqnarray}\label{on-shell_bethe_roots1}
x_k=u+\frac{\epsilon}{1-q^{\frac{1}{N}}e^{\frac{2i\pi}{N}(k-1/2)}} \,,\quad k\leq N\\ \label{on-shell_bethe_roots2}
x_k=\frac{\epsilon}{1-q^{\frac{1}{N}}e^{\frac{2i\pi}{N}(k-1/2)}} \,,\quad N<k \le 2N
\end{eqnarray}
It is very rare that interacting integrable systems admit the exact and simple expressions for the Bethe roots; we point out the similarity of our formulas to the perimeter Bethe ansatz investigated by Baxter \cite{R_J_Baxter_1987}. The price for this simplicity is, of course, the complicated limit $\epsilon\to 0$ in Eq.~\eqref{Bethe_Vector}.

The same can be said about the dual IM:
\begin{eqnarray}\label{eq:Dual_IM}
\langle \mathcal{I}|=\frac{1}{(1+q^{2})^{2N}}\left(\frac{\epsilon}{\sinh(\eta)}\right)^{N^2}\langle\mathbb{C}_{\epsilon}(\vec{x}_{\epsilon})| \Big|_{\epsilon\to 0}.
\end{eqnarray}
Before analyzing the structure of the temporal transfer matrix and the influence matrix (IM) in detail, let us first compare our construction with the quantum transfer matrix (QTM) approach introduced by Sakai in \cite{sakai2007dynamical} and further studied in \cite{pozsgay2013dynamical, piroli2017integrable} in the context of the Loschmidt echo. The QTM formalism is specifically tailored for the thermodynamic limit, relying on a particular Trotterization of the evolution operator $e^{itH}$. Consequently, the QTM involves a regularization parameter $\epsilon$, which plays the role of a discrete Trotter step $\delta t$. For each finite $\epsilon$, the QTM spectrum remains non-degenerate, and the physical dynamics is recovered only in the scaling limit where the number of time steps tends to infinity and $\epsilon\to 0$.

In contrast, the influence matrix and the integrable transfer matrix introduced in this work are always physical, directly capturing the Floquet evolution of the integrable system. The cost of this physicality is that our transfer matrix is generically non-diagonalizable, unveiling novel and previously unexplored aspects of integrability.

In the following section, we explore these properties in detail by analyzing the free-fermionic limit of the model.

\section{Free-fermionic point}\label{sec:free_fermionic_point}
{In order to analyze the structure of the IM and the Jordan blocks of the transfer matrix, let us} examine the free-fermionic point, specified by $\eta=\frac{\pi \mathrm{i}}{2}$. In this case, $\check{R}_{i,i+1}(v){=R_{i,i+1}(v) P_{i,i+1}}$ {(where $P_{i,i+1}$ is the swap of neighboring sites)} turns into:
\begin{eqnarray}
    \check{R}_{i,i+1}(v)=e^{iJ_v(\sigma_i^+\sigma_{i+1}^-+\sigma_i^-\sigma_{i+1}^+)},
\end{eqnarray}
with $J_v=\arcsin(-\tanh(v))$.

After the Jordan-Wigner transformation:
\begin{gather}\label{fermions definition}
c_i = \prod_{j < i} \sigma_j^z \sigma_i^-, \quad c_i^\dagger = \prod_{j < i} \sigma_j^z \sigma_i^+,\\
\{c_i, c_j\} = \{c_i^\dagger, c_j^\dagger\} = 0, \{c_i^\dagger, c_j\} = \delta_{i,j},
\end{gather}
$\check{R}-$operator can be identified with a quadratic fermionic operator.
\begin{eqnarray}
    \check{R}_{i,i+1}(v)=e^{-iJ_v(c_i^{\dagger}c_{i+1}+c^\dagger_{i+1}c_i)}.
\end{eqnarray}
The resulting transfer matrix from  the Eq.~\eqref{eq:Transfer_matrix} has a Gaussian form, after the projection to a sector with fixed fermionic parity. Precisely, for $q\neq 1$ we have that there are matrices $Q^\pm$, such that the transfer matrix takes the following form:
\begin{eqnarray}\label{TMatGaussForm}
\nonumber T(v)=\frac{a(v)(q^2-1)}{q^2+1}\exp\Big(\sum\limits_{i,j=1}^{4N}Q^-_{i,j} (v) c^\dagger_i c_j\Big), \ \text{odd number of particles},\\
T(v)=a(v)\exp\Big(\sum\limits_{i,j=1}^{4N}Q^+_{i,j} (v) c^\dagger_i c_j\Big), \ \text{even number of particles},
\end{eqnarray}
where $a(v)$ is the eigenvalue of the pseudovacuum: $T(v)|\Omega\rangle=a(v)|\Omega\rangle$, $a(v)=$\\
$=(\tanh(v)\tanh(u-v))^N$.

Now, in the sector with fixed parity, the conjugation of single fermionic operators by a $T(v)$ is given by:
\begin{eqnarray}\label{MMatDef}
T(v)c_i^{\dagger}T(v)^{-1}=\sum\limits_{j=1}^{4N} M^{\pm}_{j,i}(v)c_j^{\dagger},
\end{eqnarray}
where we introduced $M^\pm=\exp(Q^\pm)$, in the next section, we describe matrices $M^\pm$ in detail. {Note that we treat both $M$ and $Q$ as matrices of size $4N$, distinguishing them from corresponding operators acting on fermionic states.}
\subsection{Local Integrals}
{Matrices $M^{\pm}(v)$ commute for different values of $v$, but they are highly dense, namely all matrix elements of $M^{\pm}(v)$ are non-zero. Respectively, the transfer matrix from Eq.~\eqref{TMatGaussForm} is a highly non-local operator. In this section, we show the existence of sparse matrices $h^\pm_{l,r}$, which still commute with $M^{\pm}(v)$. These new matrices give rise to local quadratic operators $H^\pm_{l,r}$. In analogy with quantum/classical variables in Keldysh calculus (see \cite{kamenev2009keldysh} for example) it is convenient to use quantum and classical fermionic variables:} 
\begin{gather}\label{eq:q_cl_sectors}
c^{\pm,\dagger}_{i,cl}=\frac{c_i^{\dagger}\mp q^2 c_{4N+1-i}^{\dagger}}{\sqrt{1\pm q^2}}\,,\quad c^{\pm}_{i,cl}=\frac{c_i-c_{4N+1-i}}{\sqrt{1\pm q^2}}, \\
c^{\pm,\dagger}_{i,q}=\frac{c_{i}^{\dagger} + c_{4N+1-i}^{\dagger}}{\sqrt{1\pm q^{2}}}\,,\quad c^{\pm}_{i,q}=\frac{\pm q^2 c_{i}+c_{4N+1-i}}{\sqrt{1\pm q^{2}}}.
\end{gather}

{These fermionic variables are chosen to reveal the block structure of $M^\pm(v)$. Expressing $M^\pm(v)$ in terms of these variables, one finds that the terms coupling the classical and quantum sectors vanish, yielding $M^\pm(v)=M^{\pm,\mathrm{cl}}(v)\oplus M^{\pm,\mathrm{q}}(v).$}

Even more importantly, the $M^{\pm,cl/q}(v)$ entries become independent of $q$ in this new basis and coincide in sectors of different parity.

Namely, the matrix $M^{cl}(v)$ is lower-triangular, while $M^{q}(v)$ is upper-triangular, and their entries are as follows:
\begin{align}\label{MClEntries}
\nonumber M^{cl}_{2i-1,2i-1}=-\mathrm{i}\coth(u-v) \,&,\quad M^{cl}_{2i,2i}=\mathrm{i}\tanh(v),\\
\nonumber M^{cl}_{2i-1+2k,2i-1}=-\mathrm{i}\frac{\lambda^{k-1}_{cl}(v)}{\coth(v)\sinh^2(u-v)} \,&,\quad M^{cl}_{2i+2k,2i}=-\mathrm{i}\frac{\lambda^{k-1}_{cl}(v)}{\tanh(u-v)\cosh^2(v)},\\
M^{cl}_{i+2k-1,i}=-\mathrm{i}\frac{\lambda^{k-1}_{cl}(v)}{\cosh(v)\sinh(u-v)},
\end{align}
\begin{align}\label{MQEntries}
\nonumber M^{q}_{2i-1,2i-1}=-\mathrm{i}\tanh(u-v) \,&,\quad M^{q}_{2i,2i}=\mathrm{i}\coth(v),\\
\nonumber M^{q}_{2i-1,2i-1+2k}=\mathrm{i}\frac{\lambda^{k-1}_{q}(v)}{\tanh(v)\cosh^2(u-v)} \,&,\quad M^{q}_{2i,2i+2k}=\mathrm{i}\frac{\lambda^{k-1}_{q}(v)}{\coth(u-v)\sinh^2(v)},\\
M^{q}_{i,i+2k-1}=-\mathrm{i}\frac{\lambda^{k-1}_{q}(v)}{\sinh(v)\cosh(u-v)}.
\end{align}
here $\lambda_{cl}(v)=\tanh(v)\coth(u-v), \lambda_q(v)=\tanh(u-v)\coth(v)$.

In search of local integrals of motion (IOMs), let us introduce matrices $h_l,h_r$, which act locally:
\begin{align}\label{LCllEntries}
(h_l^{cl})_{2i,2i}=1, \quad (h_l^{cl})_{i+1,i}=-\frac{1}{\cosh(u)}, \quad (h_l^{cl})_{2i+1,2i-1}=1.
\end{align}
\begin{align}\label{LQlEntries}
(h_l^{q})_{2i,2i}=1, \quad (h_l^{q})_{i,i+1}=\frac{1}{\cosh(u)}, \quad (h_l^{q})_{2i-1,2i+1}=1.
\end{align}
\begin{align}\label{LClrEntries}
(h_r^{cl})_{2i-1,2i-1}=1, \quad (h_r^{cl})_{i+1,i}=\frac{1}{\cosh(u)}, \quad (h_r^{cl})_{2i+2,2i}=1.
\end{align}
\begin{align}\label{LQrEntries}
(h_r^{q})_{2i-1,2i-1}=1, \quad (h_r^{q})_{i,i+1}=-\frac{1}{\cosh(u)}, \quad (h_r^{q})_{2i,2i+2}=1,
\end{align}
and, according to the lemma below, commute with $M(v)$.
\begin{lemma}
$[M^{cl/q},h_{l/r}^{cl/q}]=0$.
\end{lemma}
\begin{proof}
Consider the matrix $h^{cl}_l$; all other cases are analogous.
\begin{align*}
    &[M^{cl},h_l^{cl}]_{2i+2k,2i} = M^{cl}_{2i+2k,2i}-\frac{1}{\cosh(u)} M^{cl}_{2i+2k,2i+1}-M^{cl}_{2i+2k,2i}
    +\frac{1}{\cosh(u)} M^{cl}_{2i+2k-1,2i}=0,\\
    &[M^{cl},h_l^{cl}]_{2i+2k-1,2i-1} = M^{cl}_{2i+2k-1,2i+1}-\frac{1}{\cosh(u)} M^{cl}_{2i+2k+1,2i}-M^{cl}_{2i+2k-3,2i-1}+\\
    &+\frac{1}{\cosh(u)} M^{cl}_{2i+2k,2i-1}=0.
\end{align*}
The other case when the indices differ by an odd number requires more care. For example,
\begin{eqnarray*}
    \nonumber [M^{cl},h_l^{cl}]_{2i+2k-2,2i-1} = M^{cl}_{2i+2k-2,2i+1}-\frac{1}{\cosh(u)} M^{cl}_{2i+2k-2,2i}-M^{cl}_{2i+2k-2,2i-1}+\\
    +\frac{1}{\cosh(u)} M^{cl}_{2i+2k-3,2i-1}.
\end{eqnarray*}
For $k=1$, the first term vanishes, leaving:
\begin{eqnarray*}
    \nonumber [M^{cl},h_l^{cl}]_{2i,2i-1} = -\frac{\mathrm{i}\tanh(v)}{\cosh(u)}+\frac{\mathrm{i}}{\cosh(v)\sinh(u-v)}-\frac{\mathrm{i}\coth(u-v)}{\cosh(u)}=\\
    \nonumber =-\mathrm{i}\frac{\sinh(v)\sinh(u-v)-\cosh(u)+\cosh(v)\cosh(u-v)}{\cosh(u)\cosh(v)\sinh(u-v)}=0.
\end{eqnarray*}
For $k>1$, all four terms contribute:
\begin{eqnarray*}
    \nonumber [M^{cl},h_l^{cl}]_{2i+2k-2,2i-1} = \mathrm{i} \lambda^{k-2}_{cl} \Big(-\frac{1}{\cosh(v)\sinh(u-v)} + \frac{1}{\cosh(u)\tanh(u-v)\cosh^2(u)}+\\
    +\frac{\tanh(v)\coth(u-v)}{\cosh(v)\sinh(u-v)}-\frac{1}{\cosh(u)\coth(v)\sinh^2(u-v)}\Big)=0.
\end{eqnarray*}
The final equality $[M^{cl},h_l^{cl}]_{2i+2k-1,2i}=0$ can be proved in the same manner.
\end{proof}
Using these matrices, one can define local and quadratic non-Hermitian `Hamiltonians' which commute with the transfer matrix:
\begin{equation}
\begin{gathered}\label{L_matrices}
H_l^{\pm,cl}=\sum\limits_{i=1}^{N}c^{\pm,\dagger}_{2i,cl} c^\pm_{2i,cl}-\sum\limits_{i=1}^{2N-1}\frac{1}{\cosh(u)}c^{\pm,\dagger}_{i+1,cl}c^\pm_{i,cl}+\sum\limits_{i=1}^{N-1}c^{\pm,\dagger}_{2i+1,cl} c^\pm_{2i-1,cl},\\
H_l^{\pm,q}=\sum\limits_{i=1}^{N}c^{\pm,\dagger}_{2i,q} c^\pm_{2i,q}+\sum\limits_{i=1}^{2N-1}\frac{1}{\cosh(u)}c^{\pm,\dagger}_{i,q}c^\pm_{i+1,q}+\sum\limits_{i=1}^{N-1}c^{\pm,\dagger}_{2i-1,q} c^\pm_{2i+1,q},\\
H_r^{\pm,cl}=\sum\limits_{i=1}^{N}c^{\pm,\dagger}_{2i-1,cl} c^\pm_{2i-1,cl}+\sum\limits_{i=1}^{2N-1}\frac{1}{\cosh(u)}c^{\pm,\dagger}_{i+1,cl}c^\pm_{i,cl}+\sum\limits_{i=1}^{N-1}c^{\pm,\dagger}_{2i+2,cl} c^\pm_{2i,cl},\\
H_r^{\pm,q}=\sum\limits_{i=1}^{N}c^{\pm,\dagger}_{2i-1,q} c^\pm_{2i-1,q}-\sum\limits_{i=1}^{2N-1}\frac{1}{\cosh(u)}c^{\pm,\dagger}_{i,q}c^\pm_{i+1,q}+\sum\limits_{i=1}^{N-1}c^{\pm,\dagger}_{2i,q} c^\pm_{2i+2,q}.
\end{gathered}
\end{equation}
{As we show in the next section, these conserved Hamiltonians share the invariant-subspace structure of the full transfer matrix, while their local form makes them simpler to analyze. We use this structure to decompose non-local matrix $M$ into invariant sectors and then determine its Jordan blocks within each sector.
} They are also closely related to those obtained in \cite{ikhlef2008staggered,bazhanov2021scaling} for alternating spin chains. 

Unfortunately, we were unable to identify local Hamiltonians commuting with the transfer matrix outside the free fermionic point. The standard approach to constructing local integrals of motion — via the logarithmic derivative of the transfer matrix evaluated at the inhomogeneity points, $H_{\alpha}^{\text{loc}} = T(v_{\alpha})^{-1} T'(v_{\alpha})$, with $v_{\alpha} \in {0, u, i\eta, u+i\eta}$ — breaks down in our case, as $T(v_{\alpha})$ is not invertible.

In the next section, we study the Jordan block structure of these Hamiltonians. As we will explain, they share the same Jordan blocks with the transfer matrix. 

{Note that the transfer matrix, as Gaussian fermionic operator preserves fermionic parity, therefore IM, as the unique eigenvector with eigenvalue one, also has fixed parity. From the Fig.~\ref{fig:IM} it can be seen that the IM is a state with even parity, because it can be constructed by applying the transfer matrix to the simple (maximally mixed) initial state belonging to the even sector. Alternatively, as we will demonstrate in the next section, this can be deduced from the analysis of eigenvalues of the transfer matrix, as one can observe that the IM is a state corresponding to the two particular fully occupied Jordan blocks. Keeping that in mind, we further restrict ourselves to an even sector only, omitting the indices $\pm$.}
\subsection{Combinatorics of Jordan blocks}

In the previous section, we introduced the action $M(v)=M^{cl}(v)\oplus M^q(v)$ of the transfer matrix on the single-particle sector, and the local Hamiltonian $h_{l/r}=h_{l/r}^{cl}\oplus h_{l/r}^q$, commuting with $M(v)$. 

The goal of the current section is to study the Jordan decomposition of the single-particle matrix $M(v)$ and its extension to the full many-body operator $T(v)$. Let us start with a Jordan decomposition of a single-particle operator:

\begin{lemma}\label{SizesOfInvSpaces}
The canonical form of the matrices $h_{l/r}^{cl/q}$ and $M^{cl/q}$ comprises exactly two Jordan blocks, each with a size of $N$.
\end{lemma}
\begin{proof}
We first notice that all given matrices are upper or lower triangular, so the entries on their diagonals are exactly the eigenvalues counted with the algebraic multiplicity. In particular, all matrices have only two distinct eigenvalues, both repeated $N$ times, which correspond to the dimension of an associated root subspace. Then, to finish the proof, it suffices to show that the geometric multiplicity of each eigenvalue is 1, i.e., there is a unique eigenvector up to scale. This fact can be checked directly.
\end{proof}

\begin{lemma}\label{LMRootSpaces}
Matrices $h_{l/r}^{cl/q}$ and $M^{cl/q}$ share common simple root subspaces of dimension $N$.
\end{lemma}
\begin{proof}
The preceding lemma guarantees that the root subspaces of all matrices are simple and have dimension $N$. Then, from the commutativity of $h_{l/r}$ and $M$, it follows that $M^{cl/q}$ acts invariantly on the root subspaces of $h_{l/r}^{cl/q}$. Finally, because the dimensions of the root subspaces coincide, the statement of this lemma holds.
\end{proof}

As we now perfectly understand the structure of the single-particle Jordan decomposition, we are ready to lift it to the many-particle sector. As we will see below, it turns out to be a non-trivial mathematical question. Luckily, the problem was partially addressed in \cite{kitahama2024jordan}, where the authors fully described the Jordan block structure of the quadratic Hamiltonian $H_k=\sum \limits_{i=1}^{M-1} \psi^{\dagger}_{i}\psi_{i-1}$. We review their results below. For now, let us outline our strategy for obtaining the Jordan block structure in our case. As a first step, we identify what we call the canonical form of the operator $T(v)$. Then, using the results of \cite{kitahama2024jordan} and one additional lemma, we derive the formula \eqref{eq:FinalMult} for the number of Jordan blocks of a given size $d$. After that, we numerically analyze this formula, showing that the number of Jordan blocks of a given size grows exponentially with the number of time steps $N$. 

Let us proceed with the canonical forms: as stated before, in the sector with even parity $T(v)$ takes the Gaussian form: {$T(v) = a(v) \,e^{\sum\limits_{i,j=1}^{4N} Q_{i,j} c_i^\dagger c_j}$, with $a(v)=(\tanh(v)\tanh(u-v))^N$}.

In this section, we are interested in the Jordan canonical form only and identify operators related by a similarity transformation $A\sim B$ if $A=OBO^{-1}$ for some invertible $O$.
We note that the matrices $M$ and $f(M)$ are similar $M\sim f(M)$ and have the same Jordan canonical structure, as long as $f$ is analytic at points $\lambda_i$ corresponding to the eigenvalues of $M$.

Let us now introduce a new set of fermions that bring $Q=\log(M)$ into its canonical form:
{\begin{equation}\label{PsiFermions}
\psi_{\alpha, i}^{\dagger}=\sum\limits_{j=1}^{4N} A^\alpha_{i,j} c_j^\dagger, \ 
\end{equation}}
{where $\alpha=1,\dots,4$ enumerates four Jordan blocks of $M$ and $\{A^\alpha\}$ are arbitrary $N \times 4N$ matrices, such that they turn $Q$ into its canonical Jordan form. In other words, we need the following property to be satisfied: $Q \psi^\dagger_{\alpha,i+1}|\Omega\rangle = \psi^\dagger_{\alpha,i}|\Omega\rangle$.  
Naturally, new fermions $\psi^\dagger_{\alpha,i}$ split the whole space of fermionic states into the tensor product of the four.}
Following that, the transfer matrix takes the form:
\begin{multline}\label{T as a product of forms}
T(v) = a(v) \,\prod\limits_{\alpha=1}^4  e^{\log{(\Lambda_\alpha)} \sum\limits_{k=1}^{N} \psi^{\dagger}_{\alpha,k} \psi_{\alpha,k} + \sum\limits_{k=1}^{N - 1} \psi^{\dagger}_{\alpha,k+1} \psi_{\alpha,k}} \sim a(v) \, e^{\sum\limits_{\alpha=1}^4\log{(\Lambda_\alpha)} \sum\limits_{k=1}^{N} \psi^{\dagger}_{\alpha,k} \psi_{\alpha,k}} \times\\
\times\big(1 + \sum\limits_{\alpha=1}^{4}\sum\limits_{k=1}^{N - 1} \psi^{\dagger}_{\alpha,k+1} \psi_{\alpha,k}\big).
\end{multline}

We call such an expression a canonical form of the transfer matrix. We see that, up to a diagonal contribution $\sum\limits_{k=1}^{N} \psi^\dagger_{\alpha,k} \psi_{\alpha,k}$, it is a sum of four canonical Hamiltonians $H_{\alpha}$:
\begin{equation}
H_{\alpha}= \sum\limits_{k=1}^{N - 1} \psi^\dagger_{\alpha,k+1} \psi_{\alpha,k}.
\end{equation}

As mentioned above, the calculation of Jordan blocks of such quadratic fermionic forms was carefully studied in \cite{kitahama2024jordan}. Let us describe their results here. Corollary 3.4 of \cite{kitahama2024jordan} states that for the nilpotent operator $H_{\alpha}$, restricted to a sector with $n$ particles, the number of Jordan blocks of a given size $d$ is given by the difference:
\begin{align}\label{DimDef}
\mathcal{N}_{n}^d=V^{[\frac{n(N-n)}{2}-\frac{d-1}{2}]}_{N,n}-V^{[\frac{n(N-n)}{2}-\frac{d+1}{2}]}_{N,n}\,,&\quad n(N-n)-d \ :\ \text{odd}\\
\mathcal{N}_{n}^d=0\ \ \ \ \ \ \ \ \ \ \ \ \ \ \ \ \ \ \ \ \ \ \ \ \ \ \ \ \ \ \ \ \ \ \ \ \ \,,&\quad n(N-n)-d  \ : \ \text{even}
\end{align}
The numbers $V^{[r]}_{M,n}$ are given in terms of the q-binomial coefficients:
\begin{eqnarray}
\begin{bmatrix} M\\
n
\end{bmatrix}_q=\frac{[M]_q!}{[M-n]_q![n]_q!}=\sum\limits_{r=0}^{n(M-n)}q^rV^{[r]}_{M,n},
\end{eqnarray}
here $[n]_q=\frac{1-q^n}{1-q}$, $[n]_q!=\prod\limits_{k=1}^n[k]_q$.

Making use of the corollary, we first decompose each of the Hamiltonians $H_{\alpha}$ into a direct sum of Jordan blocks for the fixed number of particles $n_\alpha$:
\begin{eqnarray}
H_\alpha \sim \bigoplus\limits_{d_\alpha, \ 1 \leq i_{d_\alpha} \leq \mathcal{N}^{d_\alpha}_{n_{\alpha}}} \mathcal{H}_{\alpha,d_\alpha, i_{d_\alpha}},
\end{eqnarray}
where each $\mathcal{H}_{\alpha,d_{\alpha},i_{d_\alpha}}$ is a single canonical Jordan cell with eigenvalue $0$ and size $d_\alpha$, different Jordan cells of the same size $d_{\alpha}$ are enumerated by the index $i_{d_\alpha}$. {In addition, each matrix in the right-hand side of this expression is considered in the whole space of fermions, acting nontrivially on the $\alpha$-th tensor component and by identity on the others.} Thus, we obtain the following decomposition of the transfer matrix:
\begin{equation}
T(v) \sim a(v) \, \bigoplus\limits_{n_{\alpha},d_\alpha, i_{d_\alpha}}  \big(\Lambda^{\vec{n}} \mathrm{Id} + \sum\limits_{\alpha=1}^4\mathcal{H}_{\alpha,d_\alpha, i_{d_\alpha}}\big),
\end{equation}
here $n_{\alpha}$ is the number of particles in each sector $\alpha$, and $\Lambda^{\vec{n}} = \prod\limits_{\alpha=1}^4 \Lambda_\alpha^{n_\alpha}$. We assume that we initially have a fixed number of particles in classical and quantum sectors: $n_{cl}=n_q=n_1+n_2=n_3+n_4=N$. However, if we concentrate on a fixed eigenvalue as well, this will uniquely fix all four $n_\alpha$. Despite the fact that each of the operators $\mathcal{H}_{\alpha,d_\alpha,i_{d_\alpha}}$ is already in Jordan canonical form, their sum admits a further nontrivial decomposition. To make this structure manifest, we follow \cite{kitahama2024jordan} and note that the operators $\mathcal{H}_{\alpha,d_\alpha,i_{d_\alpha}}$ can be identified with spin lowering operators $S^-_{\alpha,d_\alpha}$. These can be complemented by $S^+_{\alpha,d_\alpha}$ and $S^z_{\alpha,d_\alpha}$, forming an $\mathrm{SU}(2)$ triplet in a representation of dimension $d_\alpha$.
{Indeed, each operator $\mathcal{H}_{\alpha,d_\alpha,i_{d_\alpha}}$ acts nontrivially only on the corresponding tensor component of the Hilbert space and can be written as
\begin{equation}
\mathcal{H}_{\alpha,d_\alpha,i_{d_\alpha}}
=
\mathrm{Id}^{\otimes (\alpha-1)}
\otimes
\mathcal{H}_{\alpha,d_\alpha}
\otimes
\mathrm{Id}^{\otimes (4-\alpha)}.
\end{equation}}

{Moreover, in the Jordan basis, each operator $\mathcal{H}_{\alpha,d_\alpha}$ can be identified with the lowering operator $S^-_{\alpha,d_\alpha}$ in the $\mathrm{SU}(2)$ representation of dimension $d_\alpha$. Therefore, the sum of the four Hamiltonians takes the form
\begin{equation}
\sum\limits_{\alpha=1}^4 \mathcal{H}_{\alpha,d_\alpha,i_{d_\alpha}}
\;\sim\;
\sum\limits_{\alpha=1}^4 S^-_{\alpha,d_\alpha},
\end{equation}
i.e., it corresponds to a sum of lowering operators acting on different tensor components.} As a result, the following lemma is obtained:
\begin{lemma}
The sum of Jordan canonical matrices $\sum\limits_{\alpha=1}^4\mathcal{H}_{\alpha,d_\alpha,i_{d_\alpha}}$ admits the following decomposition:
\begin{eqnarray} \sum\limits_{\alpha=1}^4\mathcal{H}_{\alpha,d_\alpha,i_{d_\alpha}}\sim\bigoplus\limits_{D=1}^{\sum\limits_{\alpha=1}^4 d_{\alpha}-3}\bigoplus\limits_{k=1}^{C^D_{\vec{d}}} J_{D,k,\vec{i}_{d_\alpha}},
\end{eqnarray}
where $J_{D,k,\vec{i}_{d_\alpha}}$ is a Jordan cell of size $D$ and ${C^D_{\vec{d}}}$ denotes multiplicities, given by the formula:
\begin{equation}\label{CfuncDef}
{C^D_{\vec{d}}}=\frac{1}{\pi} \int_0^{2\pi} \frac{\sin(D\phi) \prod\limits_\alpha \sin(d_\alpha \phi)}{(\sin(\phi))^3} d\phi.
\end{equation}
In particular, $C^D_{\vec{d}}=0$ if $D - \sum\limits_\alpha d_\alpha$ is even.
\end{lemma}
\begin{proof}
From representation theory, it is known that the character of the irreducible representation of $\mathrm{SU}(2)$ of dimension $d$ is given by the formula $\chi_d(\phi) = \frac{\sin(d\phi)}{\sin(\phi)}$. Then, the multiplicity of the representation with dimension $D$ can be computed as the scalar product of characters:
\begin{equation*}
C^D_{\vec{d}}=\langle \chi_D|\prod\limits_\alpha \chi_{d_\alpha}\rangle = \frac{2}{\pi} \int_0^{\pi} \frac{\sin(D\phi) \prod\limits_\alpha \sin(d_\alpha \phi)}{(\sin(\phi))^5} (\sin(\phi))^2 d\phi.
\end{equation*}
In addition, it is known that the parity of dimension $D$ should coincide with the parity of $\sum\limits_\alpha d_\alpha - 1$. Otherwise, the multiplicity is zero.
\end{proof}
This lemma completes the decomposition of $T(v)$ into Jordan blocks. The final multiplicities of a Jordan block of size $D$ in the sector with fixed occupation numbers $n_\alpha$ are given by the sum:
\begin{eqnarray}\label{eq:FinalMult}
\text{Mult}^D_{N,\vec{n}}=\sum\limits_{d_{\alpha}}{C^D_{\vec{d}}}\mathcal{N}^{\vec{d}}_{\vec{n}},
\end{eqnarray}
where $\mathcal{N}^{\vec{d}}_{\vec{n}}=\prod\limits_{\alpha=1}^4\mathcal{N}^{d_{\alpha}}_{n_{\alpha}}$.

As we explain in the Appendix \eqref{app:Saddle_point}, Eq.~\eqref{eq:FinalMult} can be rewritten as a difference of two integrals:
\begin{equation}\label{FinalMultInt}
\text{Mult}^D_{N,\vec{n}}
= V^{+,D}_{N,\vec{n}} -V^{-,D}_{N,\vec{n}},
\end{equation}
where the numbers $V^{\pm,D}_{N,\vec{n}}$ are given by the following formula:
\begin{equation}\label{VIntDef}
    V^{\pm,D}_{N,\vec{n}} = \frac{1}{2 \pi \mathrm{i}} \oint \begin{bmatrix} N\\
n_1
\end{bmatrix}_q \dots \begin{bmatrix} N\\
n_4
\end{bmatrix}_q q^{-\frac{\sum\limits_\alpha n_\alpha(N-n_\alpha)+D\mp 1}{2}} dq.
\end{equation}

In the same Appendix, we argue that these integrals can be efficiently estimated by computing the large-$N$ saddle point, with the saddle point value $q^\star=e^{\frac{2\pi x^{\star}}{N+1}}$ being close to $1$: \begin{equation}\label{SaddlePointRes}
x^*_\pm=\frac{3(D\mp1)}{\pi \sum\limits_\alpha n_\alpha (N-n_\alpha)}+o(\frac{1}{N}).
\end{equation}
Substituting the saddle point into Eq.~\eqref{FinalMultInt} we obtain the approximate formula for multiplicities:
\begin{equation}\label{MultResExpand}
\text{Mult}^D_{N,\vec{n}} \approx \prod\limits_\alpha \binom{N}{n_\alpha} \Big(\frac{1}{6}\pi^{1/3} \sum\limits_\alpha n_\alpha (N-n_\alpha) (N+1)\Big)^{-3/2} D \exp\Big(-\frac{3 D^2}{2\sum\limits_\alpha n_\alpha (N-n_\alpha)(N+1)}\Big).
\end{equation}
{We call this formula the large $N$ approximation.}

\begin{figure}[H]
    \centering
    \includegraphics[trim=1.5cm 0 0 0,clip,scale=0.75]{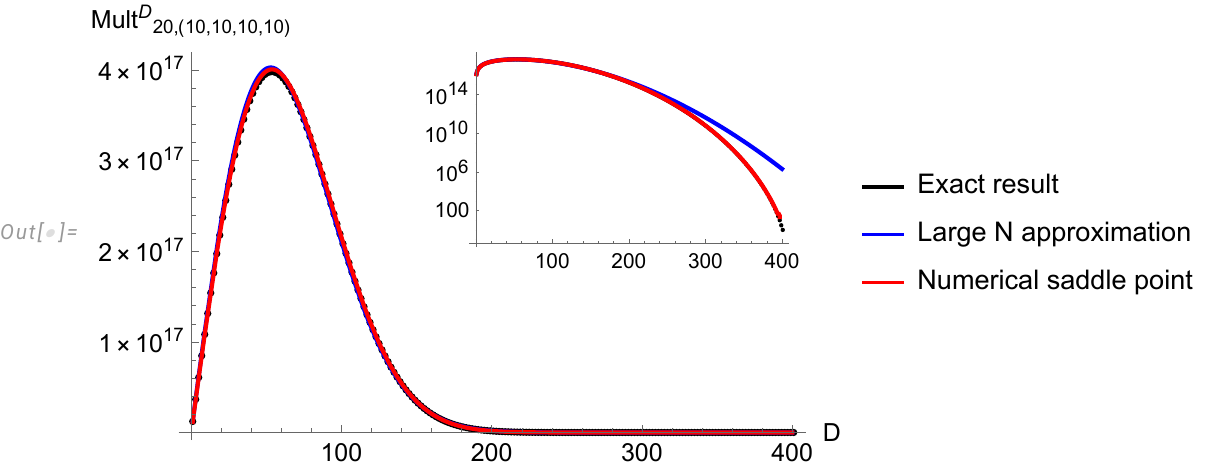}
    \caption{Comparison between the exact values of the multiplicity $\text{Mult}_{N,\vec{n}}^D$ and their approximation from the steepest descent method. The blue line is drawn according to Eq.~\eqref{MultResExpand}, while the red line is obtained by solving the saddle point equation numerically and substituting {it} into the exact q-binomial coefficients ($N=20, n_1=\dots=n_4=10$).}
    \label{fig:Dimensions}
\end{figure}


The distribution of Jordan blocks remains the same for all $v \ne 0,u$. In the latter case, the inverse of the transfer matrix from \eqref{MMatDef} is not well-defined, and our methods cannot be applied to find the degeneracy.

{The presence of exponentially many Jordan blocks of thermodynamically large size is compatible with the previously reported entanglement-barrier phenomena~\cite{lerose2023overcoming,yao2024temporal}. Indeed, under repeated application of the transfer matrix, these Jordan blocks generate a hierarchy of generalized eigenvectors, leading to highly nontrivial intermediate states during the sideways evolution. The exponential number of such blocks implies that a large number of these contributions coexist, resulting in a rapid growth of complexity and, generically, volume-law entanglement in the transient sideways dynamics.}

{At the same time, the IM corresponds to a highly structured superposition of generalized eigenstates: it fully occupies two out of the four Jordan blocks, while the remaining two are empty. This special configuration suppresses temporal entanglement compared to generic states. We analyze this structure in detail in Section~\eqref{sec:Short_range_basis}.}

\subsection{Basis of short-ranged wave functions}\label{sec:Short_range_basis}
In this section, we further analyze the spatial structure of Jordan blocks in order to reveal the entanglement structure of the IM. 

As was explained in Sec.~\ref{sec:IM_approach}, the IM is a unique eigenstate of the transfer matrix with eigenvalue 1. In the single-particle sector, the transfer matrix has four different Jordan blocks. Let us denote the corresponding wave functions that span each of the four Jordan blocks as $\Psi^{\dagger}_{cl/q,\alpha,i}$:
\begin{eqnarray}\label{Wave_func}
\Psi^{\dagger}_{cl/q,\alpha,i}=\sum\limits_{j=1}^{2N} A_{i,j}^{\alpha,cl/q} c^{\dagger}_{j,cl/q},
\end{eqnarray}
where $\alpha=0,1$ indicates a {single-particle} Jordan block with the eigenvalue $\alpha$, and the labels cl/q indicate the classical/quantum sectors introduced in Eq.~\eqref{eq:q_cl_sectors}. 
{This formula is analogous to the Eq.~\eqref{PsiFermions}, but here we chose a different way to enumerate fermionic operators, adding the index representing a classical/quantum sector.} The IM can be created by fully occupying two of the four possible Jordan blocks using the creation operators $\Psi^{\dagger}_{q,1,i}$, $\Psi^{\dagger}_{cl,0,i}$ acting on the reference vacuum state $|\Omega\rangle$:
\begin{eqnarray} \label{Creation_IM}
|\mathcal{I}\rangle\propto \prod\limits_{i=1}^{N} \Psi^{\dagger}_{q,1,i}\Psi^{\dagger}_{cl,0,i} |\Omega\rangle.
\end{eqnarray}
Indeed, it can be seen from Eq.~\eqref{TildeTM} that $T(v)|\Omega\rangle=a (v) |\Omega\rangle = (\tanh(v) \tanh(u-v))^{N} |\Omega\rangle$, and therefore:
\begin{multline}
T(v) |\mathcal{I}\rangle \propto \prod\limits_{i=1}^{N} T(v) \Psi^{\dagger}_{q,1,i} T(v)^{-1} T(v)\Psi^{\dagger}_{cl,0,i} T(v)^{-1} T(v)|\Omega\rangle
=\\
=(\Lambda_{cl,0} \Lambda_{q,1} \tanh(v) \tanh(u-v))^{N} |\mathcal{I}\rangle = |\mathcal{I}\rangle,
\end{multline}
{which uniquely defines the IM, up to a normalization factor.}
The normalization in Eq.~\eqref{Creation_IM} can be fixed by imposing the trace property defined by Eq.~\eqref{reduction}. By fixing the normalization of the wave functions as: 
\begin{eqnarray}\label{JVecNorm}
\hat{A}^{0,cl}_{N,2N-1} = \hat{A}^{1,q}_{1,2} = 1,\quad \hat{A}^{0,cl}_{i+1,k+2}=\hat{A}^{0,cl}_{i,k},\quad \hat{A}^{1,q}_{i+1,k+2}=\hat{A}^{1,q}_{i,k}
\end{eqnarray}
and requiring $|\hat{A}^{0,cl}_1\rangle,\ |\hat{A}^{1,q}_N\rangle$ to be the highest vectors, we obtain the trivial proportionality constant:
\begin{eqnarray}\label{eq:IM_Normalization}
|\mathcal{I}\rangle=\prod\limits_{i=1}^{N} \Psi^{\dagger}_{q,1,i}\Psi^{\dagger}_{cl,0,i} |\Omega\rangle,
\end{eqnarray}
where the derivation is given in Appendix \ref{sec:adjugate}.

In order to investigate the entanglement properties of the IM, {we further analyze the structure of the wave functions in Eq.~\eqref{Wave_func}, with the goal of identifying a basis in which each wave function is quasi-localized in space.} We note that the Eq.~\eqref{eq:IM_Normalization} remains invariant (up to a proportionality constant) under basis transformations inside of the linear envelope of each of the Jordan blocks:
\begin{eqnarray}
A^{\alpha,cl,q}_{i,k}\to  \hat{A}^{\alpha,cl,q}_{i,k}=\sum\limits_{j=1}^{N}C_{i,j}A^{\alpha,cl/q}_{j,k},
\end{eqnarray}
for some non-degenerate matrix $C_{i,j}$.

In the Appendix \ref{app:local_basis}, we introduce explicitly a basis in each of the Jordan blocks, which is suitable for practical computations. In particular, we provide a numerical routine for the computation of the orthogonal basis, such that the components of the wave functions obey the causality property in the sense that $i$-th wave function is supported on the first (last) $2i$ time sites, and has polynomially decaying tails:
\begin{eqnarray}\label{OrthAsym}
\begin{gathered}
\hat{A}^{0,cl}_{1,2k-1}=\hat{A}^{1,q}_{N,2N-2k+2}\approx 
\frac{B_1}{k^{3/2}}\sin (\omega(s) k - \phi_1 
)+\frac{C_1}{k^{3/2} \big(N-k+1\big)}\sin (\omega(s) k - \Phi_1) \,,\quad  k \gg 1 \\
\hat{A}^{0,cl}_{1,2k}=\hat{A}^{1,q}_{N,2N-2k+1} \approx 
\frac{B_2}{k^{3/2}}\sin (\omega(s) k -\phi_2
)+\frac{C_2}{k^{3/2}\big(N-k+1\big)}\sin (\omega(s) k - \Phi_2)\,,\quad  k \gg 1
\end{gathered}
\end{eqnarray}
where $\omega(s)=2\arcsin(s)$ and the vectors are normalized so that $\hat{A}^{0,cl}_{m,2m-1} = \hat{A}^{1,q}_{m,2m} = 1$.

\begin{figure}[H]
  \centering
  \includegraphics[scale=0.4]{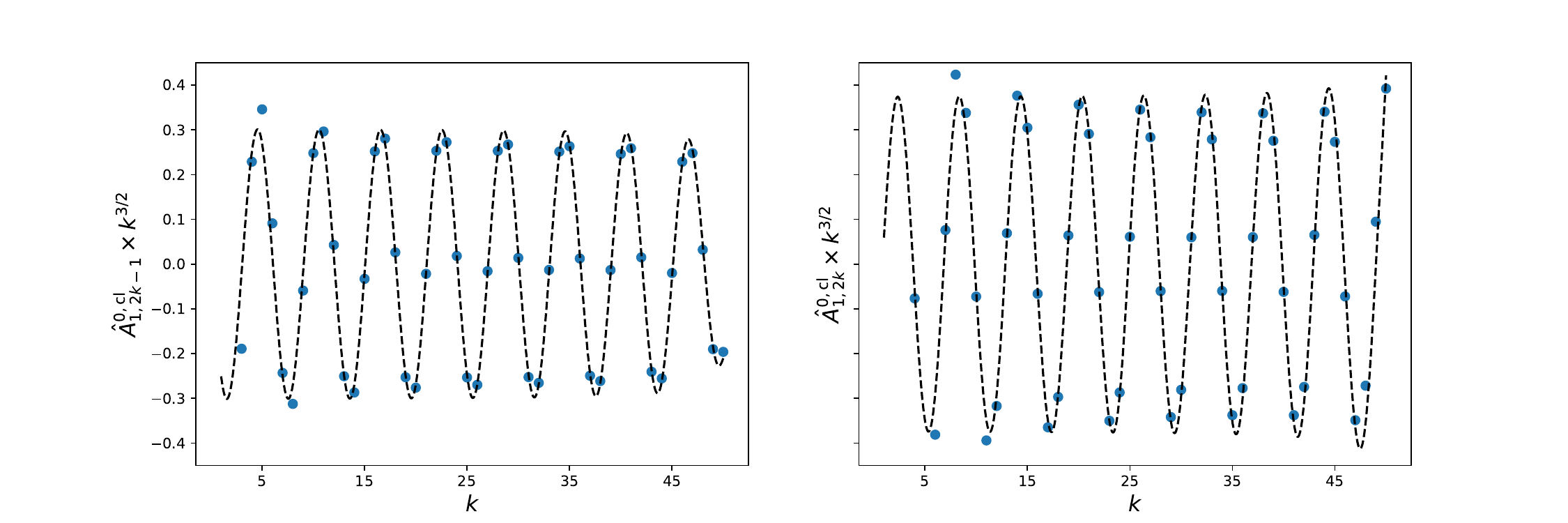}
  
  \caption{%
    Components of the highest vector in the orthogonal basis for 
    $N=50$, $s=\tfrac12$. The fitting constants are shown below:
  }
  \label{fig:JacobiOrthVector}
  
  \vspace{2ex}  
  
  \begin{tabular}{|c|c|c|c|c|c|c|c|}
    \hline
    $B_1$ & $B_2$ & $C_1$ & $C_2$ & $\phi_1$ & $\phi_2$ & $\Phi_1$ & $\Phi_2$ \\
    \hline
    0.303 & 0.371 & 0.138 & 0.210 & 3.083 & –0.880 & –0.589 & 1.735 \\
    \hline
  \end{tabular}
\end{figure}

{Orthogonal basis naturally provides a gapped Hamiltonian for which the IM is the ground state}: 

\begin{equation}
  H_{\text{gap}}=-\sum\limits_{i=1}^N (\Psi^\dagger_{cl,0,i} \Psi_{cl,0,i} + \Psi^\dagger_{q,1,i} \Psi_{q,1,i})  
\end{equation} Moreover, the decay of wave functions guarantees that such Hamiltonian is quasi-local, in agreement with a similar observation for the kicked Ising model \cite{lerose2021influence}.

\section{Conclusion}
In this article, we investigated the IM approach to the integrable Floquet dynamics. We revealed an interesting integrable structure rooted in the non-diagonalizable transfer matrix. We have conjectured an exact expression for the IM as a Bethe vector provided by Eq.~\eqref{Bethe_Vector} with exact Bethe roots given by Eqs.~(\ref{on-shell_bethe_roots1}--\ref{on-shell_bethe_roots2}).
\paragraph{Relation to QTM approach}
Our construction is similar in spirit to the so-called QTM approach introduced by Sakai in \cite{sakai2007dynamical}. While Sakai's approach works only in the continuous limit, our construction can be applied to a discretized Floquet system. This difference leads to a different organization of Bethe roots and results in novel non-diagonalizable integrable transfer matrices.
\paragraph{Free-fermionic limit to XX chain}
We also studied in detail the free-fermionic limit to the XX chain. In this case, we identified an interesting family of local Hamiltonians, commuting with the temporal transfer matrix provided by the Eqs.~\eqref{L_matrices}. 

We also managed to study the distribution of Jordan blocks given by the approximate Eq.~\eqref{MultResExpand}. This equation predicts the exponentially many blocks of thermodynamically large size, qualitatively explaining the temporal entanglement barrier effect reported earlier \cite{lerose2021influence,yao2024temporal}.

Finally, we examined the structure of the single-particle wave functions related to each of the Jordan blocks. We found that though the canonical Jordan basis leads to the unnormalizable wave functions, each invariant subspace can be spanned by the wave functions with a polynomial decay.
\paragraph{Future directions: IM approximation and correlation functions}
The polynomially decaying basis can be used to construct a matrix product state (MPS) approximation of the IM, for example, using the technique in \cite{fishman2015compression}. This approximation would provide a non-trivial tool to study the dynamics of an integrability-breaking impurity at the boundary of the free-fermionic model \cite{thoenniss2023efficient}. It is therefore natural to seek an interacting counterpart by studying perturbations away from free fermions. We expect this perturbation theory to be stable, as suggested by the logarithmic growth of the temporal entanglement of the IM in the XXZ spin chain reported in \cite{giudice2022temporal}. We plan to pursue this direction further and to develop an efficient tool for computing dynamical correlation functions in the XXZ model with broken integrability.

\section{Acknowledgments}
We are grateful to Balázs Pozsgay and Bruno Bertini for their valuable comments on a later version of this article. We also thank Alessio Lerose and Dmitry Abanin for their interest in the work and their support during its early stages. I.V. further thanks Žiga Krajnik for discussions on the IM approach to integrable systems, and Heran Wang for numerical calculations relevant to this work, which are not included in the present article.
\bibliography{MyBib}
\begin{appendix}
\section{Transfer matrix at degenerate points}\label{app:Tranfser matrix at degenerate points}
In this Appendix we study several properties of the two-site Lax operators from
Eq.~\eqref{eq:two-site_lax}, introduced at the end of Sec.~\ref{sec:YB_Transfer_matrix}.
It is sufficient to focus on three operators
$h^{\pm}(v)=L^{\pm}_{0,0}(v)$, $C^{\pm}(v)=L^{\pm}_{1,0}(v)$, and
$B^{\pm}(v)=L^{\pm}_{0,1}(v)$, since they generate the full algebra.
Explicitly,
\begin{gather}
h^+(v)=
\begin{pmatrix}
1 & 0 & 0 & 0 \\
0 & -\cot(v) & 0 & 0 \\
0 & \csc(v)\sec(v) & \tan(v) & 0 \\
0 & 0 & 0 & -1
\end{pmatrix},
\label{eq:hplus}
\\[0.6em]
C^+(v)=
\begin{pmatrix}
0 & \sec(v) & \sec(v) & 0 \\
0 & 0 & 0 & -\csc(v) \\
0 & 0 & 0 & \csc(v) \\
0 & 0 & 0 & 0
\end{pmatrix},
\label{eq:Cplus}
\\[0.6em]
B^+(v)=
\begin{pmatrix}
0 & 0 & 0 & 0 \\
\csc(v) & 0 & 0 & 0 \\
-\csc(v) & 0 & 0 & 0 \\
0 & \sec(v) & \sec(v) & 0
\end{pmatrix}.
\label{eq:Bplus}
\end{gather}
We immediately observe that the antisymmetric subspace, spanned by the single vector
\[
|a\rangle = |\uparrow\downarrow\rangle - |\downarrow\uparrow\rangle,
\]
is invariant under the action of this algebra, whereas the symmetric subspace spanned
by the remaining three basis vectors is not. For instance,
$C^+(v)\,|\downarrow\downarrow\rangle = -\csc(v)\,|a\rangle$.
Furthermore, since $B^+(v)$ and $C^+(v)$ are rank-one operators, their images do not
span the full space, for any value of the spectral parameter $v$.

A completely analogous structure appears at the opposite resonance. One finds
\begin{gather}
h^-(v)=
\begin{pmatrix}
1 & 0 & 0 & 0 \\
0 & \tan(v) & 0 & 0 \\
0 & \csc(v)\sec(v) & -\cot(v) & 0 \\
0 & 0 & 0 & -1
\end{pmatrix},
\label{eq:hminus}
\\[0.6em]
F^-(v)=
\begin{pmatrix}
0 & -\csc(v) & \csc(v) & 0 \\
0 & 0 & 0 & \sec(v) \\
0 & 0 & 0 & \sec(v) \\
0 & 0 & 0 & 0
\end{pmatrix},
\label{eq:Fminus}
\\[0.6em]
E^-(v)=
\begin{pmatrix}
0 & 0 & 0 & 0 \\
\sec(v) & 0 & 0 & 0 \\
\sec(v) & 0 & 0 & 0 \\
0 & \csc(v) & -\csc(v) & 0
\end{pmatrix}.
\label{eq:Eminus}
\end{gather}
In this case, the symmetric subspace is invariant, and the Bethe ansatz closes within it.
At the same time, the antisymmetric subspace remains orthogonal to the images of
$F^-(v)$ and $E^-(v)$.

\section{Bethe ansatz equations}\label{app:BAE}
The spectrum of the deformed transfer matrix \eqref{Tmat} can be obtained by solving the BAE:
\begin{multline}\label{eq:BAE}
\Bigg(\frac{\sinh(x_i - u -\eta)}{\sinh(x_i - u)}\frac{\sinh(x_i -\eta)}{\sinh(x_i)} \frac{\sinh(x_i +\epsilon)}{\sinh(x_i + \epsilon+\eta)} \frac{\sinh(x_i - u + \epsilon)}{\sinh(x_i - u +\eta +\epsilon)} \Bigg)^{N} =\\
=-q^{-2} \prod\limits_{j\neq i} \frac{\sinh(x_i - x_j + \eta)}{\sinh(x_i-x_j-\eta)}.
\end{multline}
We see that, formally, in the limit $\epsilon\to 0$ several factors cancel, thereby reducing the number of solutions. This indicates that the limit must be taken with care. To make these cancellations explicit, as discussed in the main text, it is convenient to rewrite the Bethe ansatz equations in the form \eqref{eq:BAEM_main}, where the $\epsilon$-dependent factors responsible for singular behavior in the limit $\epsilon \to 0$ are made explicit:
\begin{multline} \label{eq:BAEM}
\Bigg(\frac{\sinh(x_i +\epsilon)}{\sinh(x_i)}\frac{\sinh(x_i - u +\epsilon)}{\sinh(x_i - u)} \frac{\sinh(x_i -\eta)}{\sinh(x_i +\eta)} \frac{\sinh(x_i - u -\eta)}{\sinh(x_i - u +\eta)} \Bigg)^{N} =
\\= -q^{-2} \prod\limits_{j\neq i} \frac{\sinh(x_i - x_j + \eta)}{\sinh(x_i-x_j-\eta)},
\end{multline}
where we have omitted $\epsilon$ whenever it does not participate in a potential cancellation.
Let's compare \eqref{eq:BAEM} with the BAE of a spin chain on $N$ sites with $n$ magnons, arbitrary evaluation parameters $u_a$, and local sites carrying (possibly higher) spins $s_a$,
\begin{equation}
\prod\limits_{a=1}^{2N} \frac{\sinh(x_i-u_a-s_a\eta)}{\sinh(x_i-u_a+s_a \eta)}=-q^2\prod\limits_{j\neq i} \frac{\sinh(x_i - x_j + \eta)}{\sinh(x_i-x_j-\eta)}.
\end{equation}
This allows us to identify the corresponding source terms:
\begin{itemize}
    \item The source terms $\left(\frac{\sinh(x_i-\eta)}{\sinh(x_i+\eta)}\right)$ and $\left(\frac{\sinh(x_i-u-\eta)}{\sinh(x_i-u+\eta)}\right)$ correspond to spin-one sites.
\item The source terms $\left(\frac{\sinh(x_i-u+\epsilon)}{\sinh(x_i-u)}\right)$ and $\left(\frac{\sinh(x_i+\epsilon)}{\sinh(x_i)}\right)$ do not correspond to any half-integer spin. Rather, they are functions that are equal to $1$ everywhere except in the vicinity of the resonance points $x={u,0}$, respectively.

\end{itemize}

We can now classify the solutions of Eqs.~\eqref{eq:BAEM}. We first assume that all Bethe roots stay away from the resonance points, i.e. $x_k \ne u,0$. In this case the last two factors can be neglected, and the equations reduce to those of a spin-one chain on $2N$ sites, yielding $3^{2N}$ solutions.

Next, we consider configurations with $n_u$ and $n_0$ Bethe roots placed at the resonance points $u$ and $0$, respectively. The remaining $n_1=2N-n_u-n_0$ roots are generic (away from the resonances) and correspond to spin-one degrees of freedom. This motivates the following ansatz:
\begin{gather} \label{BAEsinglet1}
x^{(u)}_k=u+\epsilon \chi_k \ ,\quad \text{for} \  k\in \{1,\dots n_u\}\\ \label{BAEsinglet2}
x^{(0)}_k=\epsilon \xi_k \ , \quad \text{for} \  k\in \{1,\dots n_0\}\\ \label{BAEsinglet3}
x_k - \text{are away from the resonance points} \ , \quad \text{for}\ k\in \{1,\dots n_1\}
\end{gather}
In this case, Bethe equations split into three systems of equations:
\begin{multline}\label{eq:BAE 1}
\textrm{generic roots $x_k$}:\quad\left(\frac{\sinh(x_k-u-\eta)}{\sinh(x_k-u+\eta)}\right)^{N-n_0}\left(\frac{\sinh(x_k-\eta)}{\sinh(x_k+\eta)}\right)^{N-n_u}=\\
=-q^{-2}\prod\limits_{j=1}^{n_1}\frac{\sinh(x_k-x_j+\eta)}{\sinh(x_k-x_j-\eta)}
\end{multline}
\begin{multline}\label{eq:BAE 2}
\textrm{resonance roots}\ x^{(u)}_k:\quad\left(\frac{\chi_k-1}{\chi_k}\right)^{N}\left(\frac{\sinh(x_k-u-\eta)}{\sinh(x_k-u+\eta)}\right)^{N-n_0}\Big(-1\Big)^{N-n_u}=\\
=-q^{-2}\prod\limits_{j=1}^{n_1}\frac{\sinh(u-x_j+\eta)}{\sinh(u-x_j-\eta)}
\end{multline}
\begin{multline}\label{eq:BAE 3}
\textrm{resonance roots}\ x^{(0)}_k:\quad\left(\frac{\xi_k-1}{\xi_k}\right)^{N}\left(\frac{\sinh(x_k-\eta)}{\sinh(x_k+\eta)}\right)^{N-n_u}\Big(-1\Big)^{N-n_0}=\\
=-q^{-2}\prod\limits_{j=1}^{n_1}\frac{\sinh(x_j-\eta)}{\sinh(x_j+\eta)}
\end{multline}

The first subsystem is precisely the BAE of a spin-one chain on $2N-n_u-n_0$ sites with $2N-n_u-n_0$ magnons. If these equations can be solved (for example numerically), then the remaining two subsystems can be solved analytically.

For fixed $n_u$ and $n_0$, the total number of solutions is $3^{2N-n_u-n_0}C_{n_u}^NC_{n_0}^N$, corresponding to $3^{2N-n_u-n_0}$ solutions of a spin one chain, and $C_{n_u}^N$, $C_{n_0}^N$ solutions to Eqs.~(\ref{BAEsinglet2}--\ref{BAEsinglet3}) respectively. Summing over all possible $n_u$, $n_0$, and using an identity:
\begin{equation}
   \sum\limits_{n_u=0}^N\sum\limits_{n_0=1}^N3^{2N-n_u-n_0}C_{n_u}^NC_{n_0}^N=2^{4N}, 
\end{equation}
we reproduce all $4^{2N}$ solutions to the BAE~\eqref{eq:BAE}.
In the limit $\epsilon\to 0$, Bethe roots coalesce, producing degenerate eigenspaces of the transfer matrix. The eigenvectors of the transfer matrix are defined as a leading order contribution of the Bethe vectors from Eq.~\eqref{eq:Bvec} at $\epsilon\to 0$, and the higher vectors in the cell can be obtained {by} Taylor expansion in $\epsilon$, which induces perturbative corrections to Eqs.~(\ref{BAEsinglet1}--\ref{BAEsinglet3}). A natural expectation is that a Jordan block of size $k$ requires controlling the expansion up to order $\epsilon^{k}$; however, we do not study this systematically here and leave it for future work.
\section{Saddle point calculation of multiplicities}\label{app:Saddle_point}
In this Appendix, we analyze the large-$N$ asymptotic behavior of the summation formula from Eq.~\eqref{eq:FinalMult}. 

First, let us transform the summation into an integral representation by introducing the generating functions:
\begin{gather} 
\mathcal{N}_n(\phi) \vcentcolon= \sum\limits_{d=-\infty}^\infty \mathcal{N}_n^d q^d \ (q=e^{\mathrm{i}\phi}),\\
C^D(\phi_1,\dots,\phi_4) \vcentcolon= \sum\limits_{d_1,\dots,d_4=-\infty}^\infty C^D_{\vec{d}} q_1^{d_1} \dots q_4^{d_4} , \  \ (q_\alpha=e^{\mathrm{i}\phi_\alpha}),
\end{gather}
where we analytically continue $\mathcal{N}_n^d$ and $C^D_{d_1,\dots,d_4}$ to negative $d$ according to Eqs.~\eqref{DimDef},\eqref{CfuncDef}. These generating functions can be found explicitly:
\begin{gather}
\mathcal{N}_n(\phi) = \begin{bmatrix}N\\
n
\end{bmatrix}_{q^2} q^{-n(N-n)} (q-q^{-1}),\\
C^D(\phi_1,\dots,\phi_4)=4\pi^3\frac{\sin(D\phi_1)}{\sin^3(\phi_1)} (\delta_{2\pi}(\phi_1 - \phi_2)-\delta_{2\pi}(\phi_1 + \phi_2)) \dots (\delta_{2\pi}(\phi_1 - \phi_4)-\delta_{2\pi}(\phi_1 + \phi_4)),
\end{gather}
where $\delta_{2\pi}(x)=\sum\limits_{n\in\mathbb{Z}} \delta(x-2\pi n)$.
Then Eq.~\eqref{eq:FinalMult} turns into an integral:
\begin{multline}\label{eq:FinalMultInt_app}
\text{Mult}^D_{N,\vec{n}}=\sum\limits_{d_\alpha > 0} C^{D}_{\vec{d}} \mathcal{N}^{\vec{d}}_{\vec{n}} = \frac{1}{2^4} \sum\limits_{d_\alpha =-\infty}^\infty C^{D}_{\vec{d}} \mathcal{N}^{\vec{d}}_{\vec{n}} =\\
=\frac{1}{2^4}\frac{1}{(2\pi)^4} \int_{-\pi}^\pi \dots \int_{-\pi}^\pi C^D(\vec{\phi}) \mathcal{N}_{n_1}(-\phi_1)\dots \mathcal{N}_{n_4}(-\phi_4) d\vec{\phi}=\\
=\frac{1}{2 \pi \mathrm{i}} \oint \begin{bmatrix} N\\
n_1
\end{bmatrix}_q \dots \begin{bmatrix} N\\
n_4
\end{bmatrix}_q (q-1) q^{-\frac{\sum\limits_\alpha n_\alpha(N-n_\alpha)+D+3}{2}} dq = V^{+,D}_{N,\vec{n}} -V^{-,D}_{N,\vec{n}},
\end{multline}
where the numbers $V^{\pm,D}_{N,\vec{n}}$ are given by Eq.~\eqref{VIntDef}:

The large-$N$ limit can be treated using a saddle point, located near $q^\star\simeq 1$.
Let us analyze the q-binomial coefficient near $q=1$, expanding it in a Taylor series \cite{banerjee2017q-series}:
\begin{equation}\label{QBinomExpansion}
\log \begin{bmatrix} M\\
n
\end{bmatrix}_{q=e^{\frac{2\pi x}{M+1}}} = \log \binom{M}{n} + \frac{\pi n(M-n)}{M+1} x
+\sum\limits_{s=1}^\infty \frac{(-1)^{s+1} \zeta(2s) A^{(s)}_{M,n}}{s(2s+1)} x^{2s}, \ x\to 0,
\end{equation}
where $A^{(s)}_{M,n}=\frac{B_{2s+1}(M+1)-B_{2s+1}(M-n+1)-B_{2s+1}(n+1)}{(M+1)^{2s}}$, $B_{n}(x)$ denotes the $n$-th Bernoulli polynomial, $\zeta(n)$ is a Riemann zeta function, and  as $0 < A_{M,n}^{(s)} < M+1$, the absolute radius of convergence for the series is not less than $1$.

Then, the saddle point $x^*_\pm$ is defined as the solution to the following equation:
\begin{equation}\label{SaddlePointEq}
\sum\limits_{s=1}^\infty \frac{(-1)^{s+1} \zeta(2s) A^{(s)}_{N,\vec{n}}}{2s+1} {(x^*_\pm)}^{2s-1} = \frac{\pi (D\mp1)}{2 (N+1)},
\end{equation}
where $A^{(s)}_{N,\vec{n}} = \sum\limits_\alpha A^{(s)}_{N,n_\alpha}$.

One can iteratively solve this equation as an expansion in powers of $(N+1)^{-1}$, but we restrict ourselves to the first order only, which leads to the approximation from Eq.~\eqref{SaddlePointRes}. Substituting it into Eq.~\eqref{eq:FinalMultInt_app}, we get the following approximation:
\begin{equation}\label{VIntRes}
V^{\pm,D}_{N,\vec{n}} \approx \prod\limits_\alpha \binom{N}{n_\alpha} \frac{2\pi}{N+1} \Bigg(\frac{2\pi^3 \sum\limits_\alpha n_\alpha (N-n_\alpha)}{3(N+1)}\Bigg)^{-1/2} \exp\Big(-\frac{3(D\mp1)^2}{2\sum\limits_\alpha n_\alpha (N-n_\alpha)(N+1)}\Big).
\end{equation}
The multiplicity is given by the difference $V^{+,D}_{N,\vec{n}} - V^{-,D}_{N,\vec{n}}$. Noting that  $V^{\pm,D}_{N,\vec{n}}$ differ only by a $\mp 1$ in the exponent, we can approximate the multiplicity by further expanding the exponent $\exp\Big(-\frac{3(D\mp1)^2}{2\sum\limits_\alpha n_\alpha (N-n_\alpha)(N+1)}\Big)$ in a Taylor series, arriving at Eq.~\eqref{MultResExpand}.
\section{Exact expression for the Jordan eigenvectors}\label{sec:adjugate}
In this section, we provide an expression for the Jordan canonical vectors of matrices $h^{cl/q}_{l}$ introduced in Eqs.~(\ref{LCllEntries}--\ref{LQlEntries}), through their adjugate matrices. Note that according to lemma \ref{LMRootSpaces}, the linear envelope of these vectors provides a basis for the invariant subspaces of the matrix $M$.

For any matrix $T$, let us define the so-called adjugate matrix:
\begin{eqnarray}
\left(\text{Adj}(T)\right)_{i,j}\overset{\text{def}}{=}(-1)^{i+j}\mathcal{M}_{i,j}(T),
\end{eqnarray}
where $\mathcal{M}_{i,j}(T)$ is the $(j,i)$ minor of $T$ and $\text{Adj}(T)=\text{det}(T)T^{-1}$ if $T$ is invertible.

Now, for an arbitrary matrix $A$ consider $A(x)=\text{Adj}(x-A)$. In this notation, the following useful theorem holds:

\begin{theorem}\cite{parisse2004jordan}
Let $\lambda$ be an eigenvalue of $A$ and $n$ its algebraic multiplicity. Also, denote $A_k(x)=\frac{1}{k!} \frac{d^k}{d x^k} A(x)$. Then,
\begin{align}
(1)\quad&(A-\lambda I) A_m(\lambda) = A_{m-1}(\lambda) \ \text{for $m=1\ldots n$}, \ \text{and $(A-\lambda I) A_0(\lambda) = 0.$}\\
(2)\quad &\text{The root subspace of $A$ with eigenvalue $\lambda$ coincides with the image of $A_{n-1}(\lambda)$.}
\end{align}
\end{theorem}
This theorem implies that the vectors:
\begin{eqnarray}\label{eq:Jordan_vectors}
J_m=A_m(\lambda)|\Omega\rangle
\end{eqnarray} 
form a Jordan eigenbasis as long as $A_n(\lambda)|\Omega\rangle \ne 0$.

We apply this theorem to $A=h_l^{cl/q}$, defining $h_l^{cl/q}(x)=\mathrm{Adj}(x-h^{cl/q}_l)$. Since $h^{cl}_l$ is lower triangular and $h^{q}_l$ is upper triangular, we can immediately deduce that the highest vector of $h_l^{cl}$ is obtained by taking $\langle i|\Omega^{cl}\rangle = \delta_{i,1} $  and the highest vector of $h_l^{q}$ requires $\langle i|\Omega^{\text{q}}\rangle = \delta_{i,2N}$. 

All relevant components of the adjugate matrix take the following form:
\begin{equation}
\begin{gathered}
(h_l^{cl}(x))_{2k-1,1}=(x-1)^{N-k+1}x^{N-k}(s^2+x-1)^{k-1},\\
(h_l^{cl}(x))_{2k,1}=-s(x-1)^{N-k}x^{N-k}(s^2+x-1)^{k-1},
\end{gathered}
\end{equation}

\begin{equation}
\begin{gathered}
(h_l^{q}(x))_{2N-2k+2,2N}=\begin{cases}
s^2(x-1)^{N-k}x^{N-k+1}(s^2+x-1)^{k-2},\ k > 1\\
(x-1)^{N-1}x^{N}, \ k = 1
\end{cases}\\
(h_l^{q}(x))_{2N-2k+1,2N}=s(x-1)^{N-k}x^{N-k}(s^2+x-1)^{k-1},
\end{gathered}
\end{equation}

where $s=\text{sech}(u)$. 

Substituting these formulas into the Eq.~\eqref{eq:Jordan_vectors}, we reproduce Eqs.~\eqref{Jordan_Vector}.

As explained in Sec.~\eqref{sec:Short_range_basis}, the IM might be created from the vacuum state $|\Omega\rangle$ by the action of single-particle creation operators. Here we provide a proof that the Eq.~\eqref{eq:IM_Normalization} is compatible with the trace property of the IM from Eq.~\eqref{reduction}.
\begin{lemma}\label{lemma:WaveProdNorm}
The product of wave functions \eqref{Creation_IM} satisfies the trace property of the IM \eqref{reduction}.
\end{lemma}
\begin{proof}
First, let us recall that the wave functions generate the eigenstates of a transfer matrix $\tilde{T}$, which is derived from the original TM by conjugation with a product $\prod\limits_{i=1}^{2N} \sigma^y_{2i-1}$. In other words, the IM takes the following form:
\begin{multline}
\mathcal{I}^{\bar{s}_1,\dots,\bar{s}_{2N}}_{s_1,\dots.s_{2N}}=\Big(\prod\limits_{j=1}^{2N} \sigma^y_{2j-1} \prod\limits_{i=1}^N \Psi^\dagger_{q,1,i} \Psi^\dagger_{cl,0,i} |\Omega\rangle\Big)_{\bar{s}_{2N},\dots,\bar{s}_1,s_1,\dots,s_{2N}}=\\
=(-1)^N\prod\limits_{j=1}^N \bar{s}_{2N-2j+2} s_{2j-1}\prod\limits_{i=1}^N \Big( \Psi^\dagger_{q,1,i} \Psi^\dagger_{cl,0,i} |\Omega\rangle\Big)_{-\bar{s}_{2N},\bar{s}_{2N-1},\dots,-\bar{s}_2,\bar{s}_1,-s_1,s_2,\dots,-s_{2N-1},s_{2N}}.
\end{multline}
Then it is apparent that taking the trace of the IM involves summation over singlets $|\uparrow \downarrow\rangle - |\downarrow \uparrow\rangle$. We now proceed to prove that the trace property holds. We start by rewriting the product of wave functions in terms of original Jordan-Wigner fermions (see Eq.~\eqref{fermions definition}) explicitly:
\begin{multline}
(-1)^N\prod\limits_{i=1}^N \Psi^\dagger_{cl,0,i} \Psi^\dagger_{q,1,i} |\Omega\rangle=\\=\frac{(-1)^N}{(q^2+1)^N}\prod\limits_{i=1}^N\Big(\sum\limits_{j=2i-1}^{2N}A^{1,q}_{i,j}(c^{\dagger}_j+c^{\dagger}_{4N-j+1})\Big)
\Big(\sum\limits_{j=1}^{2i}A^{0,cl}_{i,j}(c^{\dagger}_j-q^2c^{\dagger}_{4N-j+1})\Big)|\Omega\rangle
\end{multline}
After that, one can notice that the `quantum' part of the product is traceless, while the `classical' part contributes to cancellations of $(q^2+1)^N$, since:
\begin{gather}
c^\dagger_{i_1}\ldots c^\dagger_{i_k} c^\dagger_1 c^\dagger_{i_{k+1}} \ldots c^\dagger_{i_{2N-1}} |\Omega\rangle=(-1)^k\sigma^+_1 c^\dagger_{i_1} \ldots c^\dagger_{i_{2N-1}}|\Omega\rangle,\\
c^\dagger_{i_1}\ldots c^\dagger_{i_k} c^\dagger_{4N} c^\dagger_{i_{k+1}} \ldots c^\dagger_{i_{2N-1}} |\Omega\rangle=(-1)^k\sigma^+_{4N} c^\dagger_{i_1} \ldots c^\dagger_{i_{2N-1}}|\Omega\rangle.
\end{gather}
As a result, after taking a partial trace of the IM by $s_{2N}=\bar{s}_{2N}$, one comes up with the following reduced product (here formulas ~\eqref{JVecNorm} play a crucial role):
\begin{multline}
\frac{(-1)^N}{(q^2+1)^{N-1}} (c_2^\dagger+c_{4N-1}^\dagger)\prod\limits_{i=2}^N (A^{1,q}_{i,3} c^\dagger_3+\ldots+ A^{1,q}_{i,2N} c^\dagger_{2N}+A^{1,q}_{i,2N} c^\dagger_{2N+1}+\ldots+ A^{1,q}_{i,3} c^\dagger_{4N-2}) \times\\
\times (A^{0,cl}_{i,3} c^\dagger_3+\ldots+ A^{0,cl}_{i,2N} c^\dagger_{2N}-q^2 A^{0,cl}_{i,2N} c^\dagger_{2N+1}-\ldots- q^2A^{0,cl}_{i,2} c^\dagger_{4N-2})|\Omega\rangle=\\
=(-1)^{N-1}(\sigma_2^+-\sigma_{4N-1}^+)\prod\limits_{i=2}^N\tilde{\Psi}^\dagger_{q,1,i}\tilde{\Psi}^\dagger_{cl,0,i} |\Omega\rangle,
\end{multline}
which is nothing but a product of a singlet state with the IM, corresponding to $N-1$. Thus, the proof is finished.
\end{proof}
\subsection{Local basis in Jordan cell} \label{app:local_basis}
One way to obtain the wave functions belonging to each Jordan block is to examine the free-fermionic limit of the Bethe vectors, derived in Sec.~\eqref{sec:BA}. Here we choose a different approach, to directly obtain the basis vectors, applying the method of the adjugate matrix from \cite{rubiano2024higher}. 
Let us now focus on finding appropriate wave functions $\Psi^{+,\dagger}_{cl,0,k}, \ \Psi^{+,\dagger}_{q,1,k}$. One set of the Jordan vectors can be derived using the adjugate matrix, see the Appendix \eqref{sec:adjugate}:
\begin{eqnarray}\label{Jordan_Vector}
\begin{gathered}
 \tilde{A}^{0,cl}_{m, 2k-1} = \frac{1}{(k-m)!} \ \partial_x^{k-m} \{(x-1)^{N+1-k} (s^2 - 1 + x)^{k-1}\} \big|_{x=0},\\
\tilde{A}^{0,cl}_{m, 2k} = -\frac{s}{(k-m)!} \ \partial_x^{k-m} \{(x-1)^{N-k} (s^2 - 1 + x)^{k-1}\} \big|_{x=0},\\
\tilde{A}^{1,q}_{m, 2k-1} = \frac{s}{(m-k)!} \ \partial_x^{m-k} \{x^{k-1} (s^2 - 1 + x)^{N-k}\} \big|_{x=1},\\
\tilde{A}^{1,q}_{m, 2k} = \frac{s^2}{(m-k)!} \ \partial_x^{m-k} \{x^{k} (s^2 - 1 + x)^{N-k-
1}\} \big|_{x=1},
\end{gathered}
\end{eqnarray}
where $s=\mathrm{sech}(u)$ and $k, m = 1 \dots N$.

These formulas are not practical for computations, because they contain exponentially large values. To compute the IM, we are interested in linear envelopes $\text{Span}(A^{0,cl}_1,\dots, A^{0,cl}_{N})$ and $\text{Span}(A^{1,q}_1,\dots, A^{1,q}_{N})$, rather than the Jordan blocks themselves.

A better choice of a basis, without exponentially large coefficients, is provided by the following formula:
\begin{eqnarray}\label{JacobiBasis}
\begin{gathered}
A^{0,cl}_{m, 2k-1} = -\frac{1}{(k-m)!} \ \partial_x^{k-m} \{(x-1)^{m-k+1} (s^2 - 1 + x)^{k-m}\} \big|_{x=0} = P^{(-2,0)}_{k-m} (1 - 2 s^2)\\
A^{0,cl}_{m, 2k} = \frac{s}{(k-m)!} \ \partial_x^{k-m} \{(x-1)^{m-k} (s^2 - 1 + x)^{k-m}\} \big|_{x=0}= s P^{(-1,0)}_{k-m} (1 - 2s^2),\\
A^{1,q}_{m, 2k-1} = \frac{s}{(m-k)!} \ \partial_x^{m-k} \{x^{k-m} (s^2 - 1 + x)^{m-k}\} \big|_{x=1} = s P^{(0,-1)}_{m-k} (1 - 2 s^2),\\
A^{1,q}_{m, 2k} = \frac{s^2}{(m-k)!} \ \partial_x^{m-k} \{x^{k-m+1} (s^2 - 1 + x)^{m-k-1}\} \big|_{x=1}= P^{(-1,-1)}_{m-k} (1 - 2s^2),
\end{gathered}
\end{eqnarray}
where $P^{(\alpha, \beta)}_k (z)$ is a Jacobi polynomial:
\begin{eqnarray*}
P^{(\alpha, \beta)}_k (z) = \frac{(-1)^k}{2^k k!} (1 - z)^{-\alpha} (1 + z)^{-\beta} \frac{d^k}{d z^k} \{(1 - z)^\alpha (1 + z)^\beta (1 - z^2)^k\}, \ k \geq 0\\
P^{(\alpha,\beta)}_k (z) \equiv 0, \ k < 0
\end{eqnarray*}
One can see that such defined vectors are linear combinations of previously defined Jordan vectors:
\begin{eqnarray}
\begin{gathered}\label{JacobiVector}
A^{0,cl}_{m,2k-1} = -\sum_{\tilde{m}=m}^{k} \frac{1}{(\tilde{m}-m)!} \partial_x^{\tilde{m}-m} \{(x-1)^{m-N} (s^2-1+x)^{1-m}\} \big|_{x=0} \tilde{A}^{0,cl}_{\tilde{m}, 2k-1},\\
A^{0,cl}_{m,2k} = -\sum_{\tilde{m}=m}^{k} \frac{1}{(\tilde{m}-m)!} \partial_x^{\tilde{m}-m} \{(x-1)^{m-N} (s^2-1+x)^{1-m}\} \big|_{x=0} \tilde{A}^{0,cl}_{\tilde{m}, 2k},\\
A^{1,q}_{m,2k-1} = \sum_{\tilde{m}=k}^{m} \frac{1}{(m-\tilde{m})!} \partial_x^{m-\tilde{m}} \{x^{1-m} (s^2-1+x)^{m-N}\} \big|_{x=1} \tilde{A}^{1,q}_{\tilde{m}, 2k-1},\\
A^{1,q}_{m,2k} = \sum_{\tilde{m}=k}^{m} \frac{1}{(m-\tilde{m})!} \partial_x^{m-\tilde{m}} \{x^{1-m} (s^2-1+x)^{m-N}\} \big|_{x=1} \tilde{A}^{1,q}_{\tilde{m}, 2k}.
\end{gathered}
\end{eqnarray}
These new vectors $|A^{0,cl}_n\rangle, |A^{1,q}_n\rangle$ do not provide a Jordan canonical basis, but they span the full root space.
The advantage of this new basis is a polynomial decay of the vector's components:
\begin{eqnarray}\label{VectorAsymptotes}
\begin{gathered}
A^{0,cl}_{1,2k-1} = \frac{1}{k^{1/2}} \Big(\frac{s^3}{\pi \sqrt{1 - s^2}}\Big)^{1/2} \sin ((2k - 3) \arcsin(s) - 3\pi/4) + O(k^{-3/2}) \,,\quad k\to \infty\\
A^{0,cl}_{1,2k} = \frac{1}{k^{1/2}} \Big(\frac{s^3}{\pi \sqrt{1 - s^2}}\Big)^{1/2} \sin ((2k - 2) \arcsin(s) + 3\pi/4) + O(k^{-3/2}) \,, \quad  k \to \infty\\
A^{1,q}_{N,2N-2k+1} = \frac{1}{k^{1/2}} \Bigg(\frac{s \sqrt{1 - s^2}}{\pi}\Bigg)^{1/2} \sin ((2k-2) \arcsin(s) +\pi/4) + O(k^{-3/2}) \,,\quad k\to \infty\\
A^{1,q}_{N,2N-2k+2} = \frac{1}{k^{1/2}} \Bigg(\frac{s \sqrt{1 - s^2}}{\pi}\Bigg)^{1/2} \sin ((2k-3) \arcsin(s) + 3\pi/4) + O(k^{-3/2}) \,, \quad  k \to \infty
\end{gathered}
\end{eqnarray}
\begin{figure}[H]
    \centering
    \includegraphics[scale=0.4]{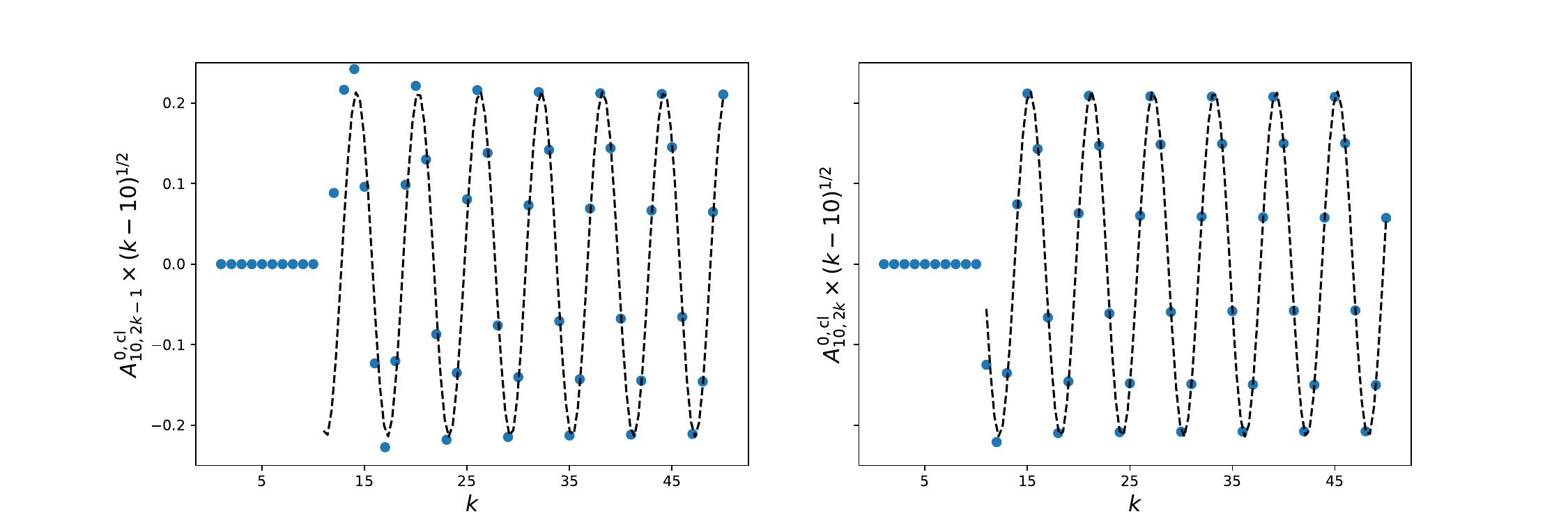}
    \includegraphics[scale=0.4]{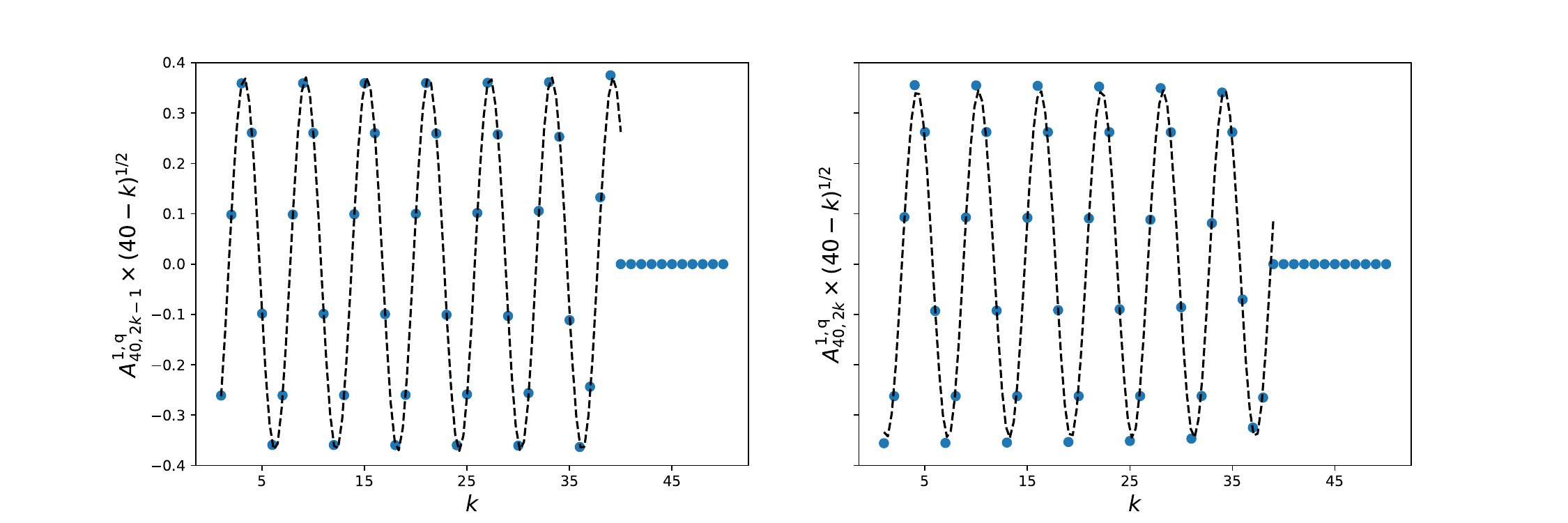}
    \caption{Components of $|A^{0,cl}_{10}\rangle$ and $|A^{1,q}_{40}\rangle$ for $N=50, \ s = \frac{1}{2}$ together with their asymptotes, which are plotted as black dashed lines.}
    \label{fig:JacobiVector}
\end{figure}

After we obtained explicit formulas for the basis components, it is straightforward to perform the Gram-Schmidt orthogonalization procedure, starting from the lowest vector $|A^{0,cl}_{N}\rangle$ (or $|A^{1,q}_1\rangle$). The resulting basis consists of vectors with the decay rate $k^{-3/2}$. We were not able to find the general formula for the components. However, numerical studies provide the conjecture for the asymptotics \eqref{OrthAsym}.
\end{appendix}
\end{document}